\documentclass[conference]{IEEEtran}

\usepackage{amsmath}
\usepackage{adjustbox}
\usepackage{multirow}
\usepackage{booktabs}
\usepackage{comment}
\usepackage{array}
\usepackage{placeins}
\usepackage{makecell}
\usepackage{subcaption}
\usepackage{xurl}
\usepackage{hyperref}
\usepackage{minted}
\usepackage[dvipsnames]{xcolor}
\usepackage{tabularx}

\newcommand{\datasetname}{\texttt{CTIRep}}
\newcommand{\subsetname}{\texttt{CTICore}}
\newcommand{\modelname}{\texttt{o3-2025-04-16}}

\usepackage[most]{tcolorbox}
\usepackage{enumitem}

\newtcolorbox{graybox}{
    colback=gray!10,       
    colframe=black,        
    width=\linewidth,      
    arc=1mm,               
    boxrule=0.5pt,         
    left=4pt,              
    right=4pt,             
    top=4pt,               
    bottom=4pt,            
}

\ifCLASSINFOpdf
\else
\fi

\begin{document}

\title{The CTI Echo Chamber: Fragmentation, Overlap, and Vendor Specificity in Twenty Years of Cyber Threat Reporting}

\author{\IEEEauthorblockN{Manuel Suarez-Roman}
	\IEEEauthorblockA{Universidad Carlos III de Madrid}
	\and
	\IEEEauthorblockN{Francesco Marchiori}
	\IEEEauthorblockA{University of Padova}
	\and
	\IEEEauthorblockN{Mauro Conti}
	\IEEEauthorblockA{University of Padova}
	\and
	\IEEEauthorblockN{Juan Tapiador}
	\IEEEauthorblockA{Universidad Carlos III de Madrid}
    }

\IEEEoverridecommandlockouts
\makeatletter\def\@IEEEpubidpullup{6.5\baselineskip}\makeatother
\IEEEpubid{\parbox{\columnwidth}{
		Network and Distributed System Security (NDSS) Symposium 2027\\
		22--26 March 2027, Seoul, Republic of Korea\\
		ISBN 978-1-970672-09-1\\  
		https://dx.doi.org/10.14722/ndss.2027.24xxxx\\
		www.ndss-symposium.org
}
\hspace{\columnsep}\makebox[\columnwidth]{}}

\maketitle

\begin{abstract}
Despite the high volume of open-source Cyber Threat Intelligence (CTI), our understanding of long-term threat actor-victim dynamics remains fragmented due to inconsistent reporting standards and the lack of structured datasets containing comprehensive analytic information. In this paper, we present a large-scale automated analysis of open-source CTI reports spanning two decades. We develop a high-precision, LLM-based pipeline to ingest and structure 16,096 reports, extracting key entities such as attributed threat actors, motivations, victims, reporting vendors, and technical indicators (IoCs and TTPs). Our analysis quantifies the evolution of CTI information density and specialization, characterizing patterns that relate specific threat actors to motivations and victim profiles. Furthermore, we perform a meta-analysis of the CTI industry itself. We identify a fragmented ecosystem of distinct silos where vendors demonstrate significant geographic and sectoral reporting biases. Our marginal coverage analysis reveals that intelligence overlap between vendors is typically low: while a few core providers may offer broad situational awareness, additional sources yield diminishing returns. Overall, our findings characterize the structural biases inherent in the CTI ecosystem, enabling practitioners and researchers to better evaluate the completeness of their intelligence sources.
\end{abstract}

\IEEEpeerreviewmaketitle

\section{Introduction}
\label{section:introduction}

Two decades of open-source Cyber Threat Intelligence (CTI) reporting have produced a substantial volume of data on threat actor activity, including their motivations, victims, behavioral traits, and technical data points about malware, infrastructure, and other tools. In recent years, the research community has built and analyzed multiple datasets on specific aspects of this ecosystem~\cite{Jin2024, Bouwman2025, saha2026kittenpandameasuringspecificity}.
Despite these efforts, our understanding of the long-term evolution of the ecosystem remains fragmented. Recent studies have emphasized the importance of longitudinal perspectives in capturing adversary dynamics~\cite{Satvat2024, Yoffe03072025, Jin2024, Yuldoshkhujaev2025}, but the field still lacks a unified framework to track these changes systematically.
One persistent problem in this domain is the lack of comprehensive datasets that contain structured and actionable analytic information
that can be systematically studied.
For example, Jin et al.~\cite{Jin2024} analyze about 6M STIX objects shared in CTI platforms between 2014 and 2023, finding that the vast majority of them contain only indicators and lack attribution and victimology data. A similar limitation applies to the dataset of more than 200M IoCs analyzed by~\cite{Bouwman2025}.
Kiavash et al.~\cite{Satvat2024} process more than 50K threat reports from 32 sources; however, they neither release the resulting dataset nor provide sufficient detail to enable reproduction of their threat report corpus.
The absence of datasets in which key information contained in CTI reports is provided in a structured format prevents the community from taking a step further. Such datasets are essential to facilitate quantitative analyses of the evolution of open-source CTI reporting and to provide a grounded characterization of the ecosystem's historical trajectory.

The emergence of Large Language Models (LLMs) with increasing capabilities has made it possible to process large quantities of CTI reports provided in natural language, typically available as HTML documents or PDF files on vendors' websites or through specialized aggregators such as Malpedia~\cite{MalpediaFraunhofer}, APT Notes~\cite{GitHubAptnotesdata}, or ORKL~\cite{githubORKL}. Prior work has focused on characterizing attacks using a variety of formalisms and techniques leveraging LLMs~\cite{Cheng2025, Tanvirul24, Yang2026, Bhusal2024} and creating structured representations like knowledge graphs~\cite{Gao2023, Wang2024,wang2024multikgmultisourcethreatintelligence, Lv2024} or STIX~\cite{Papoutsis2025, Madisetti2025, Jin2024}. Although there are concerns related to the generation of structured CTI data~\cite{Buchel25, Mezzi2025},
the rapid growth in the capabilities of LLMs makes this task increasingly affordable. For example, none of these works employ a reasoning model, nor do they consider multimodal LLM capabilities.
While recent longitudinal studies characterize APT campaigns~\cite{Yuldoshkhujaev2025}, our focus is the reporting layer itself: who reports what, how consistently strategic information is disseminated, and how vendor biases and overlaps shape the threat landscape that researchers and practitioners observe.
We address these gaps through the following questions:

\vspace{1mm}
\noindent\textbf{RQ1. Automated
extraction at scale from CTI reports.}
\emph{To what extent can LLMs reliably extract high-fidelity structured labels (campaigns, threat actors, victimology, technical indicators) from 20 years of heterogeneous, unstructured CTI text?}

\vspace{1mm}
\noindent\textbf{Contributions.}
To answer this question, we design an LLM-assisted methodology and evaluate it on a collection of 17,380
files from 
different CTI sources (\S\ref{sec:methodology}). We find that high-fidelity extraction of structured labels is viable, provided
rigorous pre- and post-processing are employed to address challenges related to the normalization of heterogeneous entities, as well as accounting for the tendency of LLMs to augment labels with external knowledge. Our validation shows a high-precision and high-recall pipeline,
yielding an overall F1-score of 94\%. The result is a dataset, \datasetname, containing 16,096 structured records from 1,926 CTI vendors, 3,867 threat actors, 12 attack motivations, 255 targeted geographies, 24 victim business sectors, 134,915 IoCs, and 935 TTPs. To our knowledge, this is the most comprehensive dataset of its nature analyzed so far
and made public. 
 
\vspace{1mm}
\noindent\textbf{RQ2. Evolution of the threat landscape and the CTI reporting activity}.
\emph{How has CTI reporting
and the strategic relationships between threat actors and victimology shifted over the last two decades?}

\vspace{1mm}
\noindent\textbf{Contributions.}
An analysis of \datasetname{} (\S\ref{sec:ecosystem}) reveals an ecosystem characterized by significant volumetric skew in terms of report types and differences in scale between technical artifacts (IoCs, TTPs) and strategic attributes (attack motivation and actor-victim relationships).
We find a strong linear correlation
($r=0.98$) between technical indicators and report volume and vendor diversity, which is not true for strategic information. Our longitudinal analysis of reporting practices identifies an inception phase (2000-2010), followed by an expansion period (2011-2019) characterized by an exponential growth in reporting, and a peak period (2020-2022) marked by sustained growth
followed by a regression toward the mean (2023-2026). We find that the threat actor landscape is extremely specialized, with over 50\% of actors reported in association with only two or fewer motivations and victim profiles (geography-sector pair) and less than 1\% of actors showing broad multi-sectoral and motivational diversity. Furthermore, the analysis of attack motivations and their relationships with victim profiles
suggests that financially motivated cybercrime (54.9\%) and espionage (33.8\%) dominate the ecosystem, although their focus is markedly different: the former targets high-wealth commercial entities in specific high-volume economies, and the latter is more geographically widespread and focuses on government and military sectors. 

\vspace{1mm}
\noindent\textbf{RQ3. Assessing vendor observational bias}.
\emph{What is the degree of reporting specificity and geographical-sectoral siloing among major CTI vendors, and how much overlap exists in their reporting of global threat campaigns?}

\vspace{1mm}
\noindent\textbf{Contributions.}
We evaluate for the first time the question of meta-intelligence and the observer effect in CTI (\S\ref{section:vendors}). We find a highly fragmented and long-tail ecosystem where 85\% of vendors are niche players, while a small set of super-vendors provide the bulk of global and multi-actor intelligence. We study quantitatively the average intelligence overlap between any two vendors and find it to be typically very low, both regarding the threat actors they track and the specific intelligence provided about a common actor. This suggests that while a few
specific vendors may suffice for broad situational awareness, deep intelligence (especially on specific threat actors) may require a highly diversified multi-vendor strategy since CTI aggregation provides diminishing returns. In general, our findings help to understand the structural biases inherent in the CTI ecosystem and provide researchers with information to evaluate the completeness of their threat intelligence sources.

\section{Related Work}
\label{sec:relatedwork}

\noindent\textbf{Automatic processing of CTI reports.}
Prior work has treated the conversion of free-form CTI reports into structured representations as a multi-step pipeline. 
Systems based on STIX, constrained prompting, or task-specific decompositions show that source grounding, ontology-aware extraction, and analyst review are essential for reliable CTI structuring~\cite{Marchiori_2023, siracusano2023timeactionautomatedanalysis, lekssays2025textactionableintelligenceautomating}. Related efforts construct CTI knowledge graphs from heterogeneous open-source reports~\cite{Gao2023,Wang2024,Cheng2025,wang2024multikgmultisourcethreatintelligence}. 
Together, these works indicate that reliable CTI structuring requires hybrid methods and careful post-processing.
Recent evaluations of LLMs for CTI extraction report mixed results. Büchel et al.~\cite{Buchel25} conclude that, due to data limitations and inherent ambiguity, LLMs do not yet outperform traditional NLP methods. 
Similarly, Mezzi et al.~\cite{Mezzi2025} find that LLMs are not sufficiently reliable to extract CTI fields from reports, exhibiting inconsistency, overconfidence, and limited improvements through adaptation. However, neither study evaluates frontier reasoning models with larger context windows, nor considers multimodal model capabilities as our work does.

\vspace{1mm}
\noindent\textbf{Analysis of the CTI ecosystem.}
Using a dataset of 6M STIX objects shared during a decade, Jin et al.~\cite{Jin2024}
discuss important limitations in the quality of available CTI datasets. Bouwman et al.~\cite{Bouwman2025} describe how adversaries often abandon IoC-related infrastructure prior to public disclosure, thus limiting its utility. Recently, Saha et al.~\cite{saha2026kittenpandameasuringspecificity} study threat actor behavioral profiles, finding that the lack of unique fingerprints complicates attribution based on public data. 
The closest work to ours is Yuldoshkhujaev et al.~\cite{Yuldoshkhujaev2025}, who analyze 1,509 CTI reports from 2014 to 2023 and derive key insights and global trends regarding past APT attacks. Our work differs in scope and analysis focus. Rather than reconstructing APT campaigns, we analyze CTI reporting as an observational ecosystem over a considerably larger corpus (over an order of magnitude larger) and focus on complementary questions such as actor-centric characterization of specialization and breadth.

\vspace{1mm}
\noindent\textbf{Vendor bias and intelligence overlap.}
Recent work has studied the way vendor visibility shapes what is perceived as the threat landscape. Vos et al.~\cite{cryptoeprint:2021/1260} present the risk of overlap and bias among CTI providers, while Bouwman et al.~\cite{10.5555/3489212.3489237} measure IoC overlap between two vendors.
Our work generalizes this question across many reporting entities and multiple intelligence types.
Ethembabaoglu et al.~\cite{AptToDisagree2026Eeten} analyze IoC overlap across seven vendors and find that shared IoCs are rare, accounting for only 1\% of the observed indicators. We study complementary dimensions, such as targeted sectors, geographies, victims, threat actors, and reporting biases, and show how vendor selection can affect broader conclusions beyond IoC-level agreement.
\section{\datasetname~Methodology}
\label{sec:methodology}
\begin{figure*}[th!]
\centering
\includegraphics[width=0.95\linewidth]{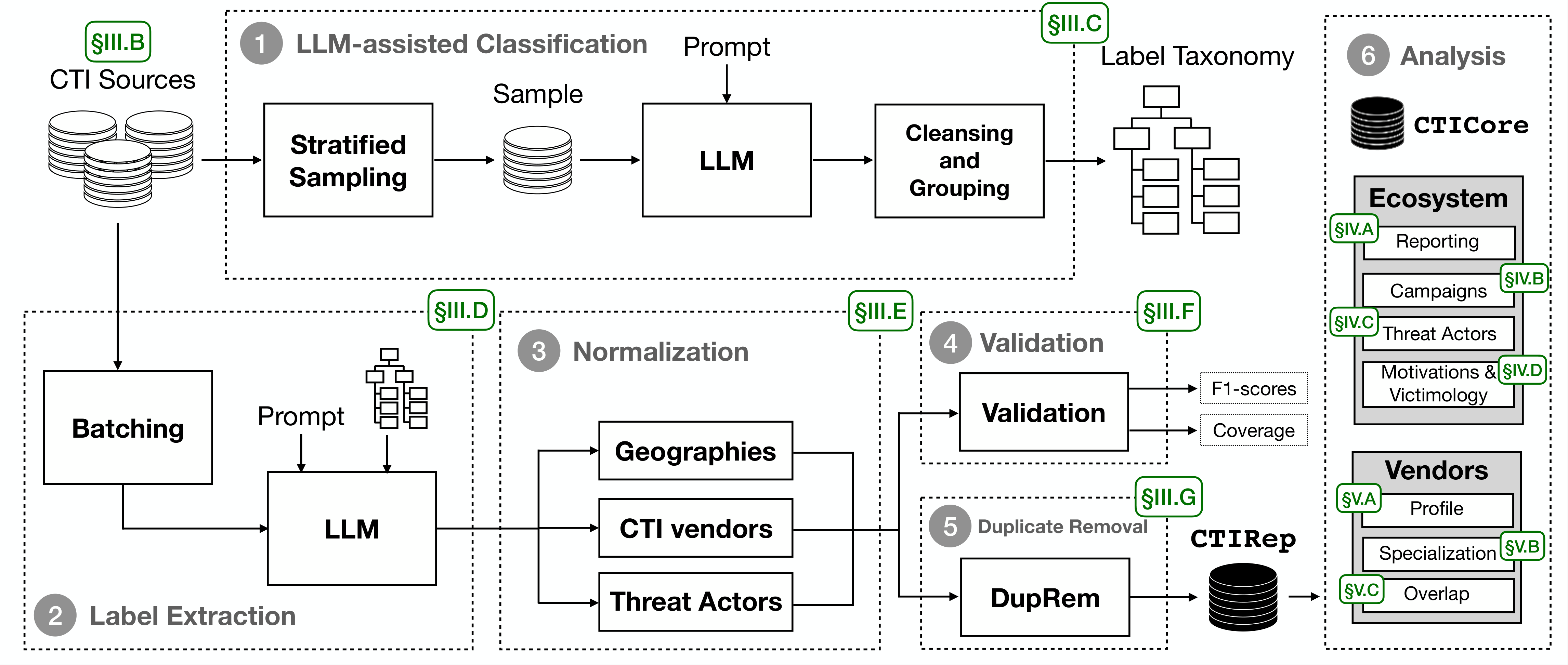}
\caption{\datasetname{} methodology and structure of the study.}
\label{fig:methodology_diagram}
\end{figure*}

This section describes the methodology used to construct \datasetname. It comprises the six phases shown in Figure~\ref{fig:methodology_diagram}. First (\S\ref{sec:source_gathering}), we collect CTI reports from multiple sources. Next (\S\ref{sec:labeling}), we construct a closed set of labels for selected fields in each report. The resulting taxonomies are used to prompt an LLM to extract structured data from each report (\S\ref{sec:processing}). The extracted features are subjected to several post-processing steps (\S\ref{sec:mappings} and \S\ref{sec:methodology:duprem}) and validation (\S\ref{sec:validation}) prior to analysis.

\subsection{Model Selection and Reproducibility}\label{sec:reproducibility}
We use several commercial LLMs offered by OpenAI at various stages of our methodology. Appendix~\ref{sec:appendix_val:othermodels} compares our choice with other alternatives, both commercial and open-source, showing that it outperforms other models. In particular, we use \texttt{o4-mini-2025-04-16} to obtain the initial taxonomy seed (\S\ref{sec:labeling}) and \modelname~to process the full report corpus (\S\ref{sec:processing}).
We decide to use a long-context window reasoning model, as they are widely recognized as better performing on complex tasks such as the one addressed in this work~\cite{10.5555/3600270.3602070, openai2025o3o4mini} and as it is a  multimodal model (i.e., it can process both text and images) and incorporate them into its reasoning process~\cite{openaiPensarImgenes}. This could overcome the limitations of LLMs in extracting CTI fields from unstructured reports discussed in \cite{Mezzi2025, Buchel25}.

\vspace{1mm}
\noindent\textbf{Reproducibility.}
One of the primary challenges when using modern LLMs for research is the variability of their outputs across runs. Moreover, because the LLM we use is proprietary, it may exhibit additional variability due to product updates. To mitigate 
this variability as much as possible and employ a reproducible methodology, we use OpenAI’s frozen model snapshots, which provide researchers with an immutable, version-locked model (including the model's weights, the tokenizer, and the model's internal configuration) that does not change among executions~\cite{openaiMakeYour}.
Other internal LLM parameters that can introduce variability in the output are the seed, the temperature, and the top-p/k parameter~\cite{unstractUnderstandingDeterministic}.
We cannot modify most of these parameters, but we are allowed to set an option to reduce variability: the reasoning effort made by the model, which we set to \texttt{high}.
To further enhance reproducibility, we include a cost analysis (see Appendix~\ref{sec:appendix:cost}) that details the technical measures employed to optimize the cost of constructing \datasetname{}.

\subsection{Sources}\label{sec:source_gathering}

We collect 17,380 files from multiple CTI report aggregators, ensuring that main CTI vendors are well represented. The main vendors in \datasetname{} are Trend Micro (630), Kaspersky (556), Google (512), Bleeping Computer (448), Palo Alto Networks (438), ESET (384), Cisco (360), Microsoft (352), Broadcom (244), Recorded Future/Insikt Group (239), and CISA (235).
All reports are provided in PDF format, although some are blog posts or web pages converted to PDF. The average report contains $2,531$ words and the mean file size is about $1.77$ MB. Overall, $96\%$ of the reports are written in English. The next most frequent languages are Russian (100), Vietnamese (77) and Korean (67). The remaining $371$ reports are written in $26$ other languages. Using \texttt{PyMuPDF}, we determine that $184$ files have a two-column layout while the majority ($10,438$) use a single column. Only $7$ reports have more than two columns, and $5,467$ use a mixed layout, with some sections written in a single column and others in multiple columns.

In \S\ref{sec:limitations}, we show how the dataset achieves sufficient coverage to render the analysis both representative and statistically meaningful.

\subsection{LLM-assisted Classification}\label{sec:labeling}

We conducted a manual inspection of a subset of reports and found that assigning certain labels can be challenging even for experienced analysts because of linguistic ambiguity and the absence of established taxonomies for CTI products. For example, \texttt{Threat Analysis} and \texttt{Threat Group Attribution} may be treated as distinct categories of report types, or may be subsumed under the broader category \texttt{Threat Actor Analysis}, which encompasses reports that provide information about a threat actor. Since one of the goals of \datasetname{} is to facilitate the study of RQ2 and RQ3, we require a finite label set for selected free-text fields to support systematic classification and analysis. To this end, we designed the following pipeline: (i) We draw a random sample of 1k reports from the entire collection. (ii) We collect the free-text values associated with report type, attack motivation, and targeted sector and, following prior work~\cite{acharya2024imitation, Galeazzi2026, Demir2026, ricaldi2026topicalshiftsdarkweb}, apply BERTopic~\cite{grootendorst2022bertopic}  separately to each dimension to induce an initial set of clusters. (iii) We manually inspect the resulting clusters, their representative terms, and the associated reports to assess their thematic coherence and identify misplaced items. (iv) We merge semantically overlapping clusters (e.g., \texttt{Cargo Transportation} and \texttt{Logistics}), split incoherent clusters when necessary, reassign misplaced items, and assign a descriptive label to each resulting cluster. (v) We organize the resulting labels into two-level taxonomies: the first level provides a coarse-grained categorization, whereas the second retains sufficient granularity to support fine-grained analysis and search.

Within the 1k report sample, the initial label space comprised 420 distinct candidate labels for report type, 80 for attack motivation, and 511 for targeted sector. After applying BERTopic and completing the manual review and consolidation steps, we obtained the final taxonomies (see Appendix~\ref{sec:appendix_taxon}): 10 first-level and 32 second-level labels for report type; 12 first-level and 25 second-level labels for attack motivation; and 24 first-level and 108 second-level labels for targeted sector. Each taxonomy includes an \texttt{Other} label to accommodate instances that do not fit any of the induced categories.

We assess the completeness of our taxonomy by manually inspecting all reports assigned \texttt{Other} for motivation, sector, and/or report type at either the 1st or 2nd taxonomy level. No report uses \texttt{Other} as a 1st-level motivation or sector. Only 11 reports use \texttt{Other} as a submotivation, and 234 as a subsector; in all cases, however, \texttt{Other} appears alongside additional submotivations or subsectors, since we allow the LLM to emit multiple labels to maximize coverage. These cases are typically triggered by broad statements such as \textit{``the malware affected every military company''}, which cause the model to select several subsectors and include \texttt{Other} to cover the remaining semantic space. For \texttt{Type of report}, 70 reports are labeled \texttt{Other}. We find 37 that correspond to empty PDFs containing only the webpage header. The remaining cases are likely due to LLM inconsistencies during processing: 31 of the 33 can be mapped to existing taxonomy categories, while the final two are glossaries or dictionaries of cybersecurity-related terms.

The taxonomies for \texttt{Motivation}, \texttt{Sector}, and \texttt{Type of Report} are needed because existing taxonomies do not fit the scope of our longitudinal study. We acknowledge that their construction requires manual effort, including a human-in-the-loop process to ensure label quality. This burden could be reduced by the availability of suitable taxonomies for these fields. Until then, after validating the completeness of our taxonomy, we open-source it to support future studies and avoid repeating this manual effort (see Appendix~\ref{sec:appendix:openscience}).

\subsection{Label Extraction}\label{sec:processing}
For each report, we prompt \modelname~to extract the fields listed in Table~\ref{tab:dataset_description} (see Appendix \ref{sec:appendix:dataset_description}). As it is a reasoning model, we craft a concise, goal-oriented prompt consistent with OpenAI’s guidance
\cite{OpenAIPlatformAdvice}.
We also leverage the role-based message prioritization supported by OpenAI models. Each request includes two prompts, one per role.

To obtain the LLM's outputs, we leverage OpenAI’s structured outputs~\cite{openaiIntroducingOutputs}. Structured outputs constrain the LLM to emit a JSON document that exactly matches a provided JSON schema. Each response contains the report metadata as a JSON document whose keys correspond to the fields listed in Table~\ref{tab:dataset_description} and whose values are arrays. For every field included in the schema, we add a brief description and examples to guide the model (e.g., \textit{``The vendor or source of the CTI report (e.g., Mandiant, ESET, etc.).''}).
Additionally, we constrain some fields with enumerations and hand-crafted rules, such as restricting the report month to the twelve month names. We also incorporate the two-level taxonomies generated in \S\ref{sec:labeling} so that the sector, motivation, and report type fields are drawn from a compact and standardized label set.
We deliberately design the output as an array, allowing \modelname~ to generate as many responses as deemed appropriate for each field, as a single report may contain information on multiple victim geographies or threat actors, may be produced by multiple vendors,
or may even span multiple report-type categories. 
The multi-label feature is both useful and common in practice: 19.5\% (3,143) of reports involve more than one geography (2.59 on average), 18.0\% (2,897) more than one threat actor (1.74 on average), and 24.7\% (3,968) more than one sector (1.53 on average).

\subsection{Normalization}\label{sec:mappings}

We post-process the output produced by the LLM to correct typographical errors and normalize values in fields that admit free text as answers: CTI vendor, threat actor, and geography. For these fields, we manually review every value assigned by the LLM across all 17,380 reports, and we perform a thorough manual consolidation as described next. We acknowledge that this effort is inherently ad hoc and difficult to automate; nonetheless, it is essential to improve data accuracy and consistency as seen in \S\ref{sec:relatedwork}.

\vspace{1mm}
\noindent\textbf{Geographies}.
One example is \texttt{Korea, Republic of}, \texttt{Republic of Korea}, and \texttt{South Korea}, all of which refer to the same country. We manually apply ad hoc unification rules for all countries and geographies in the dataset, resulting in a reduction from 364 to 255 geographic names.

\vspace{1mm}
\noindent\textbf{CTI vendors}.
CTI vendor names exhibit similar variations, such as \texttt{Malwarebytes}, \texttt{Malwarebytes labs}, \texttt{Malwarebytes threat intelligence}, and \texttt{Malwarebytes threat intelligence team}.
Further, vendor identities can change due to acquisitions, rebranding, or changes in the organizational structure. We normalize such cases to the current parent when the intelligence output is integrated. For instance, since Netscout acquired Arbor Networks, we map the following aliases to \texttt{Netscout}: \texttt{Arbor Networks}, \texttt{Arbor Networks asert}, \texttt{Netscout}, \texttt{Netscout Arbor Networks asert}, \texttt{Netscout asert}. This step reduces the set of vendors from 3,217 initial labels to 1,926.

\vspace{1mm}
\noindent\textbf{Threat actors (TAs)}.
The threat actor label is assigned the value \texttt{Unknown} when no specific name is provided in a report. Many reports list multiple aliases for the same actor, which can cause the LLM to concatenate them (e.g., \texttt{Energetic Bear (aka Crouching Yeti)}). When multiple names are detected within the same string, we retain a single canonical name according to the following hierarchy: (i) Sequential identifiers, such as \texttt{UNCXXXX}, \texttt{TAXXX}, \texttt{APTXX}, \texttt{APT-C-XX}, etc. (ii) Vendor-defined naming schemes like Microsoft~\cite{MicrosoftNames} or Palo Alto~\cite{paloaltonetworksThreatActor}; (iii) All other names.
If two names in the same string fall within the same hierarchy level, we preserve the first occurrence, which typically matches the primary designation used by the report's author.
We also unify group variants of the same TA name like adding ``Group'' or ``Gang'' at the end of the TA name, deleting the prefix ``The,'' or the inconsistent use of case and separators. 
As an example, \texttt{Cl0p}, \texttt{Clop gang}, \texttt{Clop ransomware}, \texttt{Clop ransomware gang}, \texttt{Clop ransomware group} and \texttt{Clop ransomware operators} are all mapped to the same TA name string: \texttt{Clop}.

We acknowledge that retaining only a single name when multiple aliases are provided for the same threat actor discards alias information. Alias proliferation is a well-known challenge in the CTI community~\cite{JAGS18, Roth18, Greenberg23, Poireault23}. Several prominent vendors have tried to tackle this problem by publishing curated alias lists, which we leverage to mitigate inconsistency. Specifically, we consider Microsoft and CrowdStrike’s joint naming framework~\cite{MicrosoftNames}, Palo Alto Networks Unit 42’s tracked-actor aliases~\cite{paloaltonetworksThreatActor}, Secureworks' alias catalog~\cite{secureworksCyberThreat}, MITRE's Associated groups tab~\cite{mitreGroupsMITRE} and the \textit{aka} sections in Malpedia for each of the threat actors in its database~\cite{MalpediaFraunhofer}. Because these sources differ in coverage and curation, requiring that an alias pair appear in all sources would cause an overly small unified intersection list, whereas accepting any appearance risks propagating errors due to the presence of wrong aliases in the contemplated lists. We therefore retain only alias pairs that are present in three of the five mappings, striking a balance between precision and coverage.

The resulting alias map consolidates 205 distinct names into 49 canonical TA labels. For example, the alias set for \texttt{APT29} includes \texttt{Cozy Bear}, \texttt{Midnight Blizzard}, \texttt{Nobelium}, and \texttt{UNC2452}. The alias set for \texttt{APT28} includes \texttt{Blue Athena}, \texttt{Fancy Bear}, \texttt{Fighting Ursa}, \texttt{Forest Blizzard}, \texttt{Group 74}, \texttt{Iron Twilight}, \texttt{Pawn Storm}, \texttt{Sednit}, \texttt{Snake Mackerel}, \texttt{Sofacy}, \texttt{Strontium}, \texttt{TG4127}, and \texttt{Tsar}. Finally, we map each TA label extracted from a report to its canonical name using this alias list; actors not present in the list remain unchanged. This normalization preserves a consistent and vendor-agnostic naming while minimizing loss of information from alias variability. After this step, we reduce the set of threat actors from 6,297 to 3,867 distinct labels.

As in \S\ref{sec:labeling}, this step requires some manual effort. However, because no suitable normalization rules exist for geographic entities, TAs, and vendors, we manually curated them. We discarded fully automatic normalization to maximize output quality and precision of subsequent analyses. To facilitate reproducibility and future longitudinal studies, we open-source all normalization rules used in our pipeline (see Appendix~\ref{sec:appendix:openscience}).

\subsection{Validation}\label{sec:validation}

To address the risk of hallucinations~\cite{kalai2025languagemodelshallucinate} and other errors, we assess the quality of the outputs produced by the model in the creation of \datasetname. We first process a sample of 2k reports and analyze the distribution of report types. We then select a stratified sample with a minimum of 5 reports per type,

which both aligns the precision estimate with the dataset's report type distribution and guarantees that the precision calculation captures per-type performance. 

We determine the number of reports to select based on the method used in prior studies~\cite{Yuldoshkhujaev2025, ChengW2025, Rahman2025, Froudakis2025, Zhang2025, Ahmed2024, Cheng2025}, which typically analyze between 15 and 150 reports.
Two independent experts from within the team reviewed the stratified sample, using Cohen's kappa coefficient~\cite{cohen1960coefficient} to determine inter-rater reliability.

\vspace{1mm}
\noindent\textbf{Precision, recall, and F1-score}.
We measure precision using the standard metric: the fraction of correct answers.
For reports where multiple answers are available,
if various values for the same field are produced by the LLM, the precision of that field is calculated as the number of correct answers divided by the number of answers.
We also measure the recall for those fields where the information is explicitly stated in the report but the LLM failed to provide it fully or partially. We measure the recall of a field by dividing (a) the number of correct answers provided by the model by (b) the total number of answers (correct and missed).
Finally, we compute the F1-score using the standard definition.
The overall quality of field extraction across all report types is high, with an overall F1-score of 93.8\% (94.6\% precision and 93.4\% recall). Multiple fields achieve a perfect score, indicating that the model is a strong metadata extractor for several fields, yet significant variance exists within specific categories and fields. \texttt{Academic (ACA)} and \texttt{IoCs, TTPs \& Indicators (ITI)} stand out as the main underperformers, with F1-scores of 76.9\% and 86.6\%, respectively. This is likely due to the fact that these reports are the least similar to other \datasetname{} report types: their content often lacks information for several fields, such as motivation, sector, or geography, and in some cases they do not describe an actual attack. Looking at the individual fields, the lowest scores are concentrated in the motivation/submotivation (87.8\% and 86.6\%) fields, as well as the IoCs (88.6\%), suggesting that these are more subjective or harder to extract accurately than other metadata such as the report year or title. In the case of academic reports, the model struggles with identifying the victim sector/subsector and threat actor, indicating a potential content-related challenge specific to that report category.
Detailed tables with precision, recall and F1-score values for each field and type of report are provided in Appendix~\ref{sec:appendix_val:precrec}.

Due to the severe class imbalance across report types, the unadjusted validation mean is not 
representative of corpus-level performance, since the stratified validation sample 
over-represents minority report types. To correct for this sampling design, we computed post-stratified weighted estimates using the true distribution of report types in the complete corpus ($N = 16,096$). Using report-level precision and recall values, with F1 computed at the field level and then averaged per report, the population-adjusted estimated macro-F1 is 93.0\%. Uncertainty was estimated using a stratified bootstrap over the manually validated reports, yielding a 95\% confidence interval of [91.0\%, 94.8\%]. The corresponding post-stratified precision and recall are 94.5\% and 96.6\%, respectively.

\vspace{1mm}
\noindent\textbf{Output stability.}
We assess the stability and robustness of model outputs. We repeatedly submit a random subset of $10$ reports to \modelname~and measure stability by requesting the processing of each of the 10 files 10 times. We then calculate the precision for each field and run.
To quantify differences in the precision values across runs, we compute the Shannon entropy ($H$ in Table~\ref{tab:f1_selected_reports}) for each field. We find that, on average, 13 of the 16 fields have identical precision across runs, while the remaining non-identical fields show only two distinct outcomes (typically either perfectly correct or entirely incorrect). The report title, year, month, vendor, TA, and IoCs always yield identical precision across runs. The fields with more variability are not equally distributed across reports, nor is the number of distinct per-run precision values per field. The overall entropy is less than $0.05$, indicating that the model rarely changes outputs across repeated runs.
Additionally, some fields exhibit virtually no variability, whereas others vary more (most notably the campaign duration and the TTPs).

\vspace{1mm}
\noindent\textbf{Beyond compliance.}
We manually inspect the 100 reports and observe several instances where the model's capabilities, despite being 
accurate, force us to consider them errors. In some cases, the model correctly identifies the report title, although it is absent from the file due to a PDF conversion error. This likely occurred because the report contains the source URL, which the model used to infer the title.
Similarly, in other cases the model identifies the date the report was created based on links found on images embedded in the report. Another interesting case is the extraction of IoCs and TTPs in reports that do not explicitly list any.
We find that the model extracts the hash of the reported malware sample(s) based on VT links (as each VT URL contains the artifact's SHA-256 hash~\cite{virustotalFiles}) and includes the corresponding IoC in the output.
Another case concerns a report describing an attack against Indian embassies in Saudi Arabia and Kazakhstan. The model correctly identifies the victim geography as India, arguing that embassies are attributed to the sending state rather than the host country.

We mark all these cases as errors in our evaluation, as the information present in the report and the answer provided by the model differ. Although the answer is technically correct, our focus is on the model's faithfulness to the source document, not on its ability to enrich it with external knowledge. Even when doing so, the overall performance remains strong, achieving an F1-score of 93.8\%.

\subsection{Removing Duplicates}
\label{sec:methodology:duprem}

Based on the high precision of \modelname{} for extracting the report title and year, we perform a 
deduplication stage.
We group the 17,380 reports by the tuple (title, year). This flags 2,103 files that collapse into 978 title groups. The remaining 15,277 reports have unique titles. Not all items sharing the same title and year are duplicates: some reports are explicitly multi-part; appendices and executive summaries may be published as separate documents; and some are updates to an earlier release. Such cases are retained.
In total, we identify and remove 943 duplicates. Common causes include: page orientation or layout differences (portrait vs. landscape, one, two or three columns), presence or absence of embedded images (we retain the version with images), variations in web footers (e.g., site owner information, inclusion/exclusion of blog comments), minor aesthetic revisions across crawl times, and superficial branding changes (like vendor or team naming). We also remove, among others, two clusters of 7 and 11 files consisting of anti-scraping artifacts rather than actual CTI information.

\begin{table}[t!]
\centering
\caption{Distribution of entities in \datasetname{} and \subsetname.}
\label{tab:cti_datasets}
\resizebox{0.95\linewidth}{!}{%

\begin{tabular}{lrrrrrrrr}
\toprule
 Type & \# Rep. & \# Vend. & \# TAs & \# Geo. & \# Sec. & \# Mot. & \# IoCs & \# TTPs \\
\midrule
MVA & 7,188 & 1,007 & 1,565 & 186 & 24 & 12 & 67,855 & 727 \\
CA & 2,932 & 444 & 1,228 & 200 & 24 & 9 & 40,983 & 643 \\
OMC & 2,536 & 738 & 920 & 155 & 24 & 10 & 4,632 & 270 \\
TAA & 2,305 & 427 & 1,597 & 205 & 24 & 7 & 25,933 & 621 \\
TLT & 1,185 & 273 & 1,192 & 146 & 23 & 9 & 5,342 & 400 \\
IHF & 357 & 169 & 175 & 50 & 22 & 8 & 3,094 & 242 \\
ACA & 226 & 177 & 143 & 56 & 19 & 9 & 581 & 145 \\
ITI & 202 & 95 & 129 & 37 & 21 & 7 & 5,379 & 127 \\
CPL & 50 & 26 & 44 & 10 & 15 & 5 & 246 & 1 \\
OTH & 70 & 39 & 7 & 4 & 4 & 2 & 10 & 4 \\
\midrule
\datasetname & 16,096 & 1,926 & 3,867 & 255 & 24 & 12 & 134,915 & 935 \\
\subsetname & 11,862 & 1,287 & 2,996 & 243 & 24 & 12 & 124,999 & 893 \\
\bottomrule
\end{tabular}
}
\end{table}

Table~\ref{tab:cti_datasets} reports the number of entities in the resulting dataset (\datasetname).
The results are disaggregated by report type and include a derived dataset, \subsetname{},
composed of reports of type \texttt{Malware and Vulnerability Analysis (MVA)}, \texttt{Campaign Analysis (CA)}, \texttt{Threat Actor Analysis (TAA)} and \texttt{ITI}, which is described in the following section.

\begin{graybox}
\textbf{Takeaway.} High-precision LLM-assisted labeling for large-scale CTI datasets is viable, provided rigorous pre- and post-processing are employed to normalize heterogeneous entities such as vendor, threat actor, and victimology data. Furthermore, researchers must account for the tendency of models to augment labels with external knowledge, a behavior that may enhance or compromise dataset integrity depending on the research objective.
\end{graybox}

\section{CTI Ecosystem Overview}
\label{sec:ecosystem}

To answer RQ2, we analyze \datasetname{} from three complementary perspectives. First, we examine the types of information reported over time and the reporting entities (\S\ref{section:ecosystem:reporting}). Then, we introduce \subsetname{}, a subset of \datasetname{} specifically built to support cyberattack campaign analysis (\S\ref{section:ecosystem:attackevol}). Finally, we analyze the threat actors involved in the campaigns (\S\ref{section:ecosystem:tas}) and we examine their motivations and characterize victims (\S\ref{section:ecosystem:motvic}).

\subsection{CTI Reporting Types and Evolution}
\label{section:ecosystem:reporting}

The data in Table~\ref{tab:cti_datasets} suggest an ecosystem characterized by significant volumetric skew and differences in scale between technical artifacts and strategic attributes.

\vspace{1mm}
\noindent\textbf{Information density and specialization.}
IoCs represent the primary volume of \datasetname, totaling 134,915 and appearing in 66.9\% (10,772) of the reports, which is approximately 35 times the number of the next largest entity class (threat actor at 3,867). High-level entities such as sectors and motivations remain approximately constant across the aggregate dataset, which is not surprising since we force the model to extract them from a finite set of categorical labels. We also observe a non-uniform distribution of entities across report types, indicating varying levels of information. \texttt{MVA} is the most prolific category, contributing 7,188 reports and 67,855 IoCs, and accounting for roughly 50.3\% of all indicators in the repository. The highest technical density is, unsurprisingly, found in reports of type \texttt{ITI}, which contribute an average of $\approx 26.6$ IoCs per report, compared to \texttt{MVA}'s $\approx 9.4$.
\texttt{Incident Handling \& Forensics (IHF)} exhibits the highest tactical density with {0.68} TTPs per report,
outperforming \texttt{TAA} (0.27) and \texttt{CA} (0.22). Despite its high volume (2,536 reports), \texttt{Outreach, Media \& Communications (OMC)} contains only 1.82 IoCs and 0.11 TTPs per report, suggesting a high degree of redundancy or less structured data extraction. {Only 17.2\% (2,766) list TTPs, yielding 935 distinct TTPs out of the 992 described in the latest MITRE ATT\&CK versions (including ATLAS)~\cite{mitre_attack_v18}.}

Analysis of the sector and motivation distributions reveals both some overlap and completeness of the report types. Four categories of reports (\texttt{TAA}, \texttt{CA}, \texttt{MVA}, and \texttt{OMC}) cover 100\% of the 24 identified sectors. \texttt{MVA} is the only category that achieves 100\% coverage of all 12 motivation types. \texttt{Compliance, Policy and Legal (CPL)} and \texttt{Other (OTH)} represent the most isolated categories, with sector coverage as low as {16.7}\%. 
Reports in these categories likely focus on highly specific, non-overlapping legal or administrative domains.

\vspace{1mm}
\noindent\textbf{Cross-category correlation analysis.}
We perform a multivariate correlation analysis and find a strong linear coupling between report volume and technical indicators ($r = 0.98$
for IoCs) and vendor diversity ({$r = 0.98$}). This suggests that technical data grows linearly with dataset expansion. However, the correlation between report volume and high-level strategic entities like sectors ({$r = 0.42$}) and motivations ({$r = 0.72$}) is much lower.
This might suggest that a small subset of reports is sufficient to saturate the strategic knowledge space, whereas technical artifacts require continuous, high-volume reporting.

\begin{figure*}[ht!]
\centering
\includegraphics[width=0.95\linewidth]{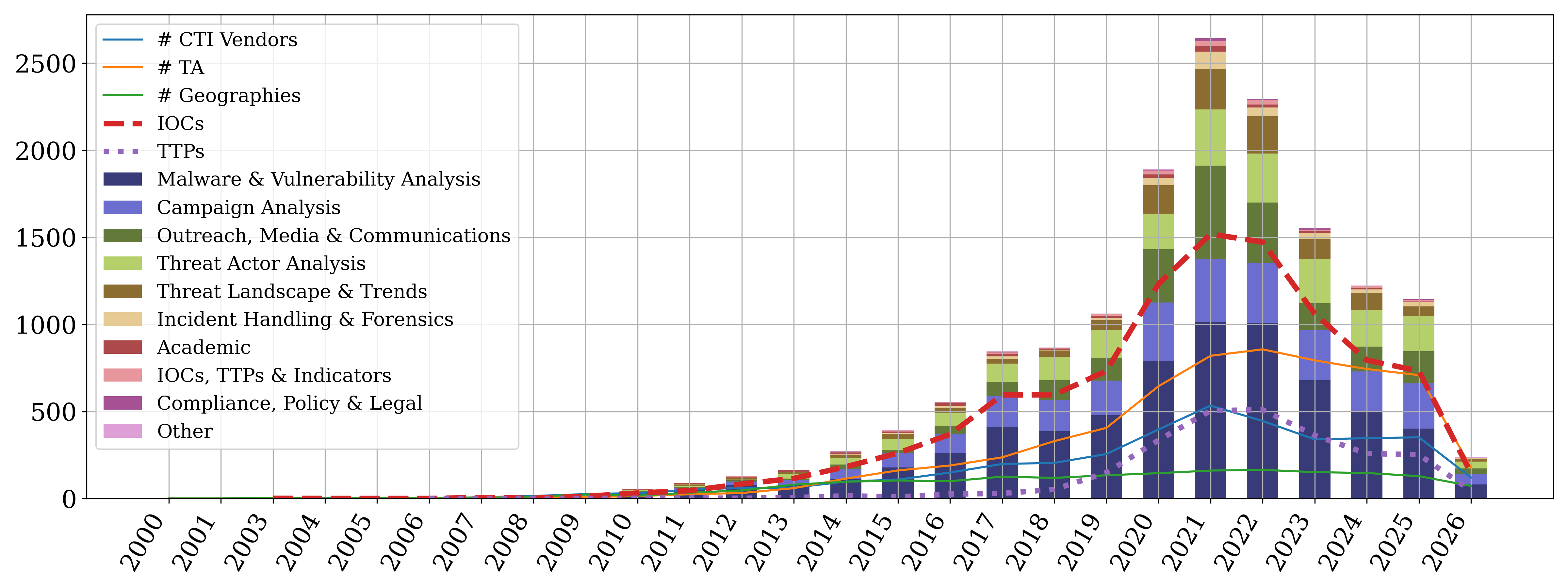}
\caption{Temporal evolution of the volume, diversity, and distribution of report types present in \datasetname.}
\label{fig:stacked_report_types}
\end{figure*}

\vspace{1mm}
\noindent\textbf{Longitudinal analysis.}
We examine the evolution over time of the volume of reports and their type distribution (see Figure \ref{fig:stacked_report_types}). We also track four variables each year: the number of vendors publishing reports, the number of threat actors mentioned, the number of different victim geographies, and the number of reports that include IoCs or TTPs. 
Publication volume increases over time at a faster-than-linear growth, peaking in 2021 with 2,520 documents.
For the remaining years, we can identify four distinct phases of growth: (i) An inception phase (2000–2010), in which the report volume is negligible, representing a nascent period of CTI activity; (ii) an expansion period (2011–2019), marked by a steady exponential growth in reporting; (iii) a peak activity period (2020–2022) characterized by a significant surge in volume; and (iv) a decline toward the volumetric levels observed in the second phase (2023-2026). We cannot determine the exact causes of this decline; however, it may be related to observations in prior work suggesting that, during the pandemic period, cybercrime (and, consequently, the volume of cyberattack reports) increased significantly~\cite{LALLIE2021102248, interpol2020cybercrime, Hoheisel2023}.

The breakdown of report types reveals shifts in the focus of the CTI industry. \texttt{MVA} and \texttt{CA} consistently form the foundation of the dataset across all years. From 2017 onward, categories such as \texttt{OMC} and \texttt{TAA} grow significantly, suggesting an expansion toward behavioral and public-facing intelligence. A Jensen-Shannon (JS) homogeneity test was conducted to assess differences in the distribution of report types across years. The observed mean JS distance was 0.12, indicating that yearly distributions, particularly from 2019 onward, remain substantially closer to the pooled composition.

The growth of IoCs, TTPs, threat actors, geographies, and vendors follows different patterns. The volume of IoCs is strongly correlated with the total number of reports, an observation consistent with the findings above about their prevalence in the dataset. TTPs have a minimal presence in the dataset until 2018, followed by a sharp climb suggesting wide adoption by the industry.
The number of vendors shows moderate growth, plateauing at approximately 400-500
vendors during the 2020–{2025} period. Finally, the number of geographies is the most stable metric, reaching a saturation point around 2015.

\begin{graybox}
\textbf{Takeaway.}
CTI is characterized by a massive volumetric skew where technical indicators (IoCs) scale linearly with reporting, while strategic insights like motivations and sectors saturate quickly. After an exponential growth phase starting in 2011, the ecosystem matured post-2017 by expanding its focus from malware analysis toward behavioral intelligence (TTPs) and public-facing narratives. CTI reporting volume then increased sharply in the post-pandemic period, followed by a regression toward the mean.
\end{graybox}

\subsection{Campaigns over the Years}
\label{section:ecosystem:attackevol}

Table \ref{tab:top_years} shows the distribution of the number of attack campaigns reported over a two-decade period, along with the top threat actors, motivations, and targeted geographies and sector per year. 
For this analysis, we restrict the corpus to reports of type \texttt{TAA}, \texttt{CA}, \texttt{MVA} and \texttt{ITI}. We will refer to this subset, composed of {11,862} reports, as \subsetname. This selection is designed to only consider documents that reliably link each threat actor to its associated information and activity. Including \texttt{TLT} or other types of report would conflate data between multiple actors mentioned in the same document.
\begin{table}[t!]
\caption{Volume of reported campaigns over the years and top threat actors, targeted geographies, sectors and motivations.}

\label{tab:top_years}
\centering
\resizebox{0.93\linewidth}{!}{
\begin{tabular}{lrllll}
\toprule
Year & Vol. & \#1 Geo. & \#1 TA & \#1 Sector & \#1 Motiv. \\
\midrule
2010 & 42 & USA (6) & Zeus (3) & IPS (11) & FC (23) \\
2011 & 61 & USA (12) & Night dragon (2) & EUNR (12) & EIC (25) \\
2012 & 101 & China (11) & Luckycat (4) & FBI (19) & EIC (39) \\
2013 & 130 & USA (28) & Red october (8) & GPA (51) & EIC (77) \\
2014 & 201 & USA (53) & Turla (10) & GPA (73) & EIC (128) \\
2015 & 305 & USA (66) & Fancybear (10) & GPA (95) & EIC (167) \\
2016 & 421 & USA (80) & Fancybear (26) & GPA (130) & EIC (195) \\
2017 & 655 & USA (81) & Lazarus (26) & FBI (132) & FC (322) \\
2018 & 656 & USA (80) & Fancybear (28) & GPA (177) & FC (331) \\
2019 & 799 & USA (98) & TA505 (25) & GPA (181) & FC (455) \\
2020 & 1,273 & USA (189) & Lazarus (40) & GPA (278) & FC (739) \\
2021 & 1,640 & USA (263) & Revil (52) & ITTC (322) & FC (1016) \\
2022 & 1,583 & USA (191) & Conti (52) & GPA (320) & FC (948) \\
2023 & 1,135 & USA (123) & Lazarus (49) & GPA (240) & FC (665) \\
2024 & 867 & USA (106) & Lazarus (21) & GPA (212) & FC (460) \\
2025 & 795 & USA (117) & Lazarus (32) & ITTC (204) & FC (485) \\
2026 & 162 & USA (23) & Fancybear (8) & ITTC (43) & FC (90) \\
\bottomrule
\end{tabular}
}
\end{table}

Geographically, the US receives the highest volume of attacks ({1,603}), followed by Ukraine ({568}), South Korea ({527}), India ({513}), and Russia ({459}). Motivations have undergone a fundamental transition: while the early 2010s were dominated by \texttt{Espionage \& Intelligence Collection (EIC)}, the landscape shifted heavily toward \texttt{Financial/Cybercrime (FC)} starting in 2017, coinciding with the professionalization and expansion of ransomware groups.
The primary victim business sector is, in nearly every year, \texttt{Government \& Public Administration (GPA)}, suggesting a persistent focus on state-related targets. The extracted data indicates a reliance on \texttt{MVA} as the main type of CTI documentation, being the most frequent type of report for every year.

Our findings are mostly consistent with the recent analysis by Yuldoshkhujaev et al.~\cite{Yuldoshkhujaev2025}, although their dataset is significantly smaller (1,509 reports covering the 2014-2023 period). We also find that the most frequent targets are government and defense sectors in the US, reinforcing the premise that state-level infrastructure is a primary victim of cyberattacks. There is also significant overlap in terms of (a) threat actor dominance, particularly with \texttt{Turla}, \texttt{Fancy Bear/APT28} and \texttt{Lazarus} during the 2016–2021 window; and (b) frequent geographies other than the US, with South Korea (2018 and 2023) and Ukraine (2022) as notable cases since they were among the most attacked geographies in these years.

\subsection{Threat Actor Characterization}
\label{section:ecosystem:tas}

\begin{figure}[t]
\centering
\includegraphics[width=0.95\linewidth]{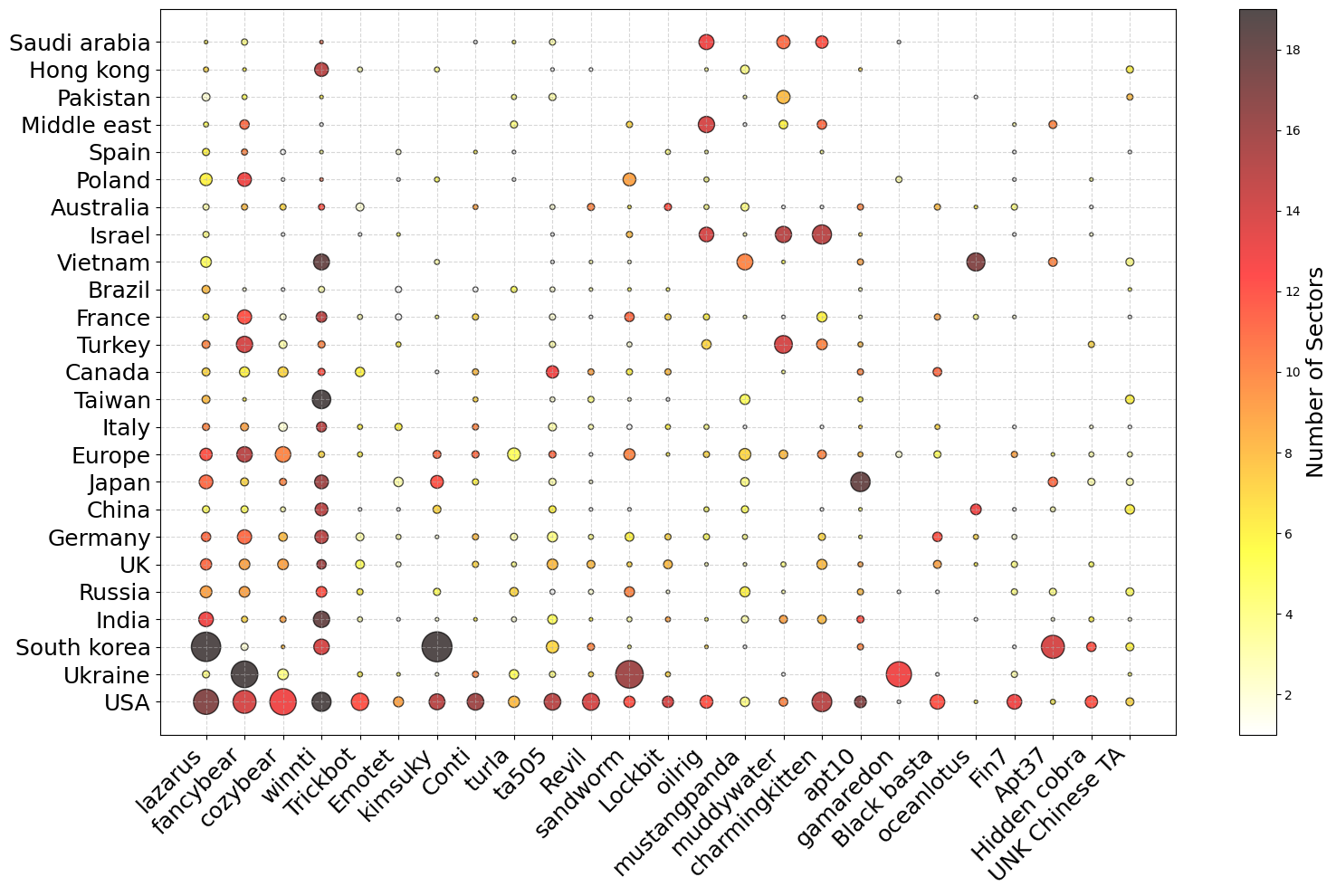}
\caption{Distribution of the top 25 geographies (bottom to top) attacked by the top 25 threat actors (left to right). The bubble size and color represent no. of reports and no. different business sectors of the victims, respectively.}
\label{fig:country_ta_reports_sectors}
\end{figure}

We characterize the activity of the threat actors present in \subsetname, focusing on the actor–victim relations that constitute the vertical axis of the Diamond Model~\cite{caltagirone2013diamond}. We discuss the geographies and business sectors targeted by each threat actor (see Figure~\ref{fig:country_ta_reports_sectors}) and the motivations of these attacks.
We do not distinguish between conventionally defined APTs and non-APT actors, as the boundary between them is increasingly questioned~\cite{crowdstrike2020Global, paloaltonetworksExtortionRansomware, welivesecurityStatealignedGroups, microsoft2025Microsoft, googleCybercrimeMultifaceted}, and CTI reporting often links state-aligned and criminal-like activity~\cite{10.5555/3766078.3766191, Prasad2025}.
Table~\ref{tab:TA_characterization} details, for each actor, the number of distinct first- and second-level motivations and sectors, and victim geographies, along with the actor's total number of appearances across reports. We find that {58\%} of actors appear only once in the dataset; however, a single report may contain several motivations, geographies, and sectors.

\begin{table}[t]
\caption{Percentage (and count) of threat actors having different number of motivations, attacked sectors and geographies of the victims. \texttt{Unknown} values are not counted.}
\label{tab:TA_characterization}
\centering
\resizebox{0.9\linewidth}{!}{%
\begin{tabular}{lllll}
\toprule
\# Threat Actor & 1 & 2-4 & 5-9 & $\ge 10$ \\
\midrule
Motivation & 80.8\% (2,423) & 17.0\% (510) & 0.4\% (12) & 0.0\% (0) \\
Submotivation & 52.0\% (1,558) & 44.0\% (1,319) & 2.2\% (65) & 0.1\% (3) \\
Sector & 29.5\% (885) & 32.1\% (962) & 18.4\% (552) & 5.0\% (149) \\
Subsector & 23.4\% (700) & 31.7\% (951) & 20.1\% (603) & 9.8\% (294) \\
Geography & 19.1\% (573) & 22.3\% (668) & 14.3\% (430) & 11.8\% (354) \\
Appearances & 58.0\% (1,739) & 28.1\% (841) & 7.8\% (234) & 6.1\% (182) \\
\bottomrule
\end{tabular}
}
\end{table}

\vspace{1mm}
\noindent\textbf{Motivational and geographical specialization.}
Most threat actors are observed with a single primary motivation and submotivation, and are linked to a single victim geography. By contrast, the sectoral breadth is mixed: It is approximately as common for an actor to target a single industry as it is to target more than one but fewer than five. Actors with diversity of motivations are rare: only {0.4\% (12)} have 5 or more distinct motivations and {5.0\% (149)} target ten or more sectors. However, {11.8\% (354)} are linked to 10 or more distinct victim geographies. 
48.7\% (1,460) of threat actors are only linked to two or fewer motivations, sectors, and geographies.
Examples of specialized actors include \texttt{APT10}, with {30} attacks against victims in Japan spanning {17} sectors, and \texttt{Sandworm}, for which {75\% (60)} of attacks target Ukrainian victims.

\vspace{1mm}
\noindent\textbf{Exceptions.}
Some notable TAs constitute an exception to the global pattern of specialization. One example is \texttt{Fancy Bear}, with the largest set of distinct motivations (6) and geographies {(87)}. \texttt{Winnti} shows the widest variety of victim business sectors (24) and subsectors ({63}) in {73} geographies. \texttt{Lazarus Group}, like \texttt{Fancy Bear}, has the highest number of unique motivations (6) and ranks first in the number of submotivations {(14)}. It also ranks {second} in number of geographies ({75}) and sectors attacked ({20}).

\begin{graybox}
\textbf{Takeaway.}
The threat actor landscape is defined by extreme specialization and niche operations, where over 49\% of actors focus on only two or fewer motivations, sectors and geographies, and {58}\% appear only once in the dataset. While geographic expansion is relatively common, true sectoral and motivational diversity is a rare trait found in less than 1\% of the threat actors.
\end{graybox}

\subsection{Motivations and Victimology Patterns}
\label{section:ecosystem:motvic}

\begin{figure}[t!]
\centering
\includegraphics[width=\linewidth]{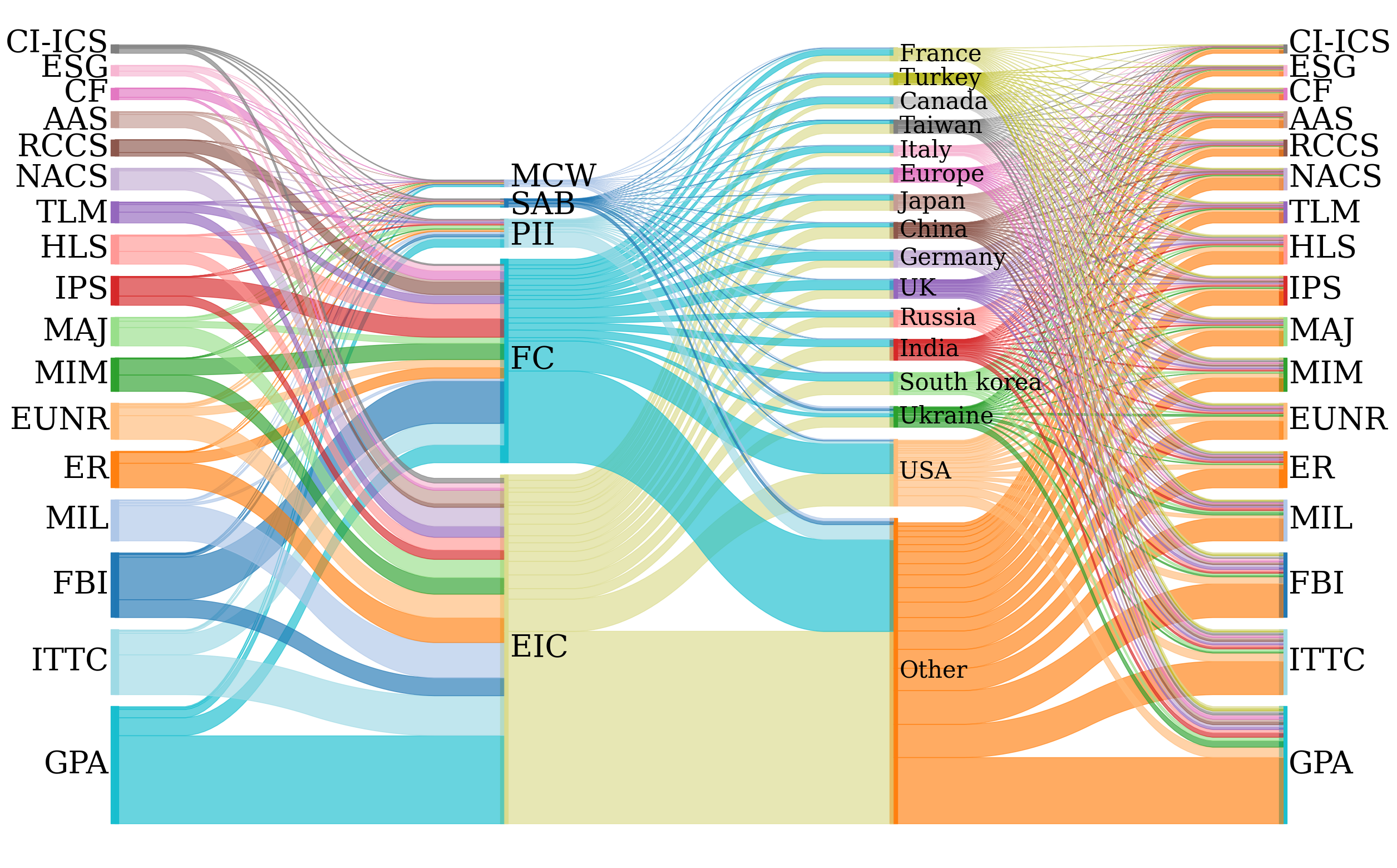}
\caption{Distribution of the top victim business sectors, the top attack motivations and the most attacked geographies. Acronyms are described in Table \ref{table:taxonomies}.}
\label{fig:sankey}
\end{figure}

We study the potential relationships between motivations, targeted geographies and sectors to characterize attack patterns to assess whether attacks received by each geography are similarly motivated and whether they affect industry sectors uniformly. We base our analysis on Figure~\ref{fig:sankey}, which shows, on the left, the relationship between the motivations behind cyberattacks and the targeted business sectors. On the right, it illustrates how these motivations flow into specific geographies and then back into the affected domestic sectors.
A ranking of the top five targeted geographies, disaggregated by motivation and sector, is provided in Table~\ref{tab:top_countries} (Appendix~\ref{sec:appendix:topvictims}).

\vspace{1mm}
\noindent\textbf{Motivations vs. business sectors.}
The distribution of motivations is not uniform, with \texttt{FC} accounting for 54.9\% (6,513) of attacks, \texttt{EIC} for {33.8\% (4,015)}, and the remaining motivations together for {11.3\% (1,334)}. \texttt{EIC} acts as the primary driver for attacks against \texttt{GPA} {(80.3\%)}, \texttt{Information Technology, Telecommunications \& Cybersecurity (ITTC)} {(49.5\%)} and \texttt{Military (Mil)} {(92.9\%)}. This motivation is particularly linked to the US and a broad set of other nations, suggesting that state-sponsored or high-level intelligence gathering is a global phenomenon.
In contrast, \texttt{FC} shows a more diverse spread across commercial sectors such as \texttt{FBI} {(82.1\%)}, and \texttt{Retail, Commerce \& Consumer Services (RCCS)} {(89.6\%)}, with a significant concentration of these financially motivated flows also directed toward the US ({56.1\%} of the total attacks against this country), the UK ({63.1\%}) and Italy (73.4\%). An interesting outlier is Ukraine, where the
attacks motivated by \texttt{Sabotage (S)} and by \texttt{Military \& Cyber Warfare (MCW)} are {18.1\%} and {10.6\%} respectively, far above the cross-geography average of 2.8\% and 0.8\%.

\vspace{1mm}
\noindent\textbf{Geographies vs. motivations.}
The median per-geography share for \texttt{EIC} is {65.1}\%, whereas \texttt{FC}'s median is {33.2}\%. These figures suggest that espionage-motivated activity is widespread across many geographies, while \texttt{FC} is concentrated in a subset of geographies with high attack volumes. For example, Brazil ({85.1}\%) and Spain ({80.8}\%) rank {16}th and {21}st overall by total attacks, yet both show high \texttt{FC} shares. In contrast, the top three \texttt{EIC}-heavy geographies --- North Korea (100.0\%), Brunei (100\%) and Cambodia (98.2\%)
--- rank {110}th, {120}th and {58}th, 
respectively. Both \texttt{FC} and \texttt{EIC} show high standard deviation (22.8 and 22.4, respectively), confirming that their distributions are far from uniform across geographies.
All other motivations display more homogeneous behavior, each with an Inter-Quartile Range below {7.6}\%. Notable outliers include Ukraine (18.1\%), Albania (14.3\%) and Saudi Arabia (11.2\%) 
for \texttt{Sabotage}; Latvia (36.4\%), Estonia (34.8\%) and Lithuania (33.3\%) 
for \texttt{Political \& Ideological Influence (PII)}; Tunisia (18.2\%) for \texttt{Malware Propagation (MP)}; and {Ukraine (10.6\%), Central Asia (9.3\%) and Latvia (9.1\%) 
for \texttt{MCW}.

\vspace{1mm}
\noindent\textbf{Business sectors vs. geographies.}
Business sectors targeted by cyberattacks are unevenly distributed. The top three most targeted sectors are \texttt{GPA}  {(21.0\%, 2,493 reports)}, \texttt{ITTC} {(16.3\%, 1,934 reports)}, and \texttt{FBI} {(14.0\%, 1,663 reports)}. Variation within sectors is substantial: \texttt{GPA} shows the largest IQR ({30.5}\%), the highest standard deviation ({19.2}\%), and no outliers are detected, indicating a heterogeneous per-geography distribution of \texttt{GPA}-targeted attacks.

\begin{graybox}
\textbf{Takeaway.}
The global threat landscape is dominated by financially motivated cybercrime ({54.9}\%) and espionage ({33.8}\%). While the former targets high-wealth commercial hubs in high-volume economies, the latter focuses on government and military sectors and is more geographically widespread, often being the primary threat for smaller or developing nations. In some regional outliers, like Ukraine and the Baltic states, geopolitical friction shifts the motive from profit or intelligence toward sabotage and ideological warfare.
\end{graybox}

\section{Vendor Analysis}
\label{section:vendors}
To address RQ3, we analyze vendor behavior across three axes: publication volume and typology (\S\ref{section:vendor:activity}); TA tracking specialization (\S\ref{section:vendor:specialization}); and coverage overlap across the ecosystem (\S\ref{section:vendor:overlap}).

\vspace{1mm}
\noindent\textbf{A note on vendors.}
In \datasetname, a `vendor' is defined as the reporting entity of a report. Vendors primarily comprise commercial CTI firms, but also include government agencies (e.g., CISA) and specialized cybersecurity media outlets (e.g., Bleeping Computer). In our analysis, we identify vendors using generic labels ($V_i$) instead of their official names. This is not an attempt at anonymization, as identifying the vendors is immediate given the open-source nature of the dataset. Instead, our approach aims to highlight ecosystem-level findings rather than the practices of specific vendors.

\begin{figure}[t]
\centering
\includegraphics[width=\linewidth]{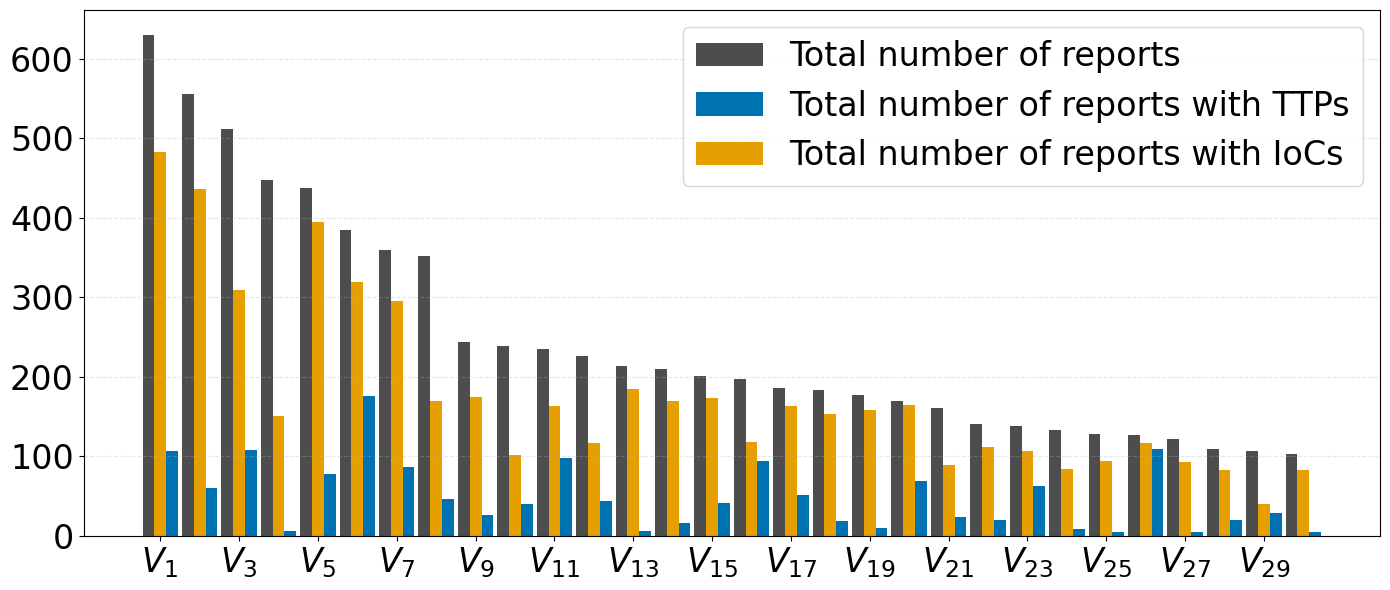}
\caption{Number of reports (total count, reports including TTPs, and reports including IoCs) for the top 30 vendors.}
\label{fig:vendor_TTP_IOC}
\end{figure}

\subsection{Vendor Publication Volume and Typology}
\label{section:vendor:activity}

\noindent\textbf{Published volume.}
Figure~\ref{fig:vendor_TTP_IOC} shows the number of published reports for the top 30 vendors out of the {1,926} identified in \datasetname{}. For each vendor, we also show the number of reports containing IoCs and TTPs.
The distribution of reports per vendor is highly skewed, with the top 50 vendors accounting for {55.6}\% ({8,948}) of the total volume. 

\vspace{1mm}
\noindent\textbf{Published content.}
We find that IoC reporting is far more widespread than TTP reporting: {123} vendors include IoCs in more than 75\% of their reports, and {19} vendors include IoCs in every report. In contrast, {1,479} vendors never publish any TTP, but only {776} never publish IoCs. Among vendors with more than 10 reports, only 6
include TTPs in more than 75\% of their reports.
Vendor practices also differ in the number of distinct TAs tracked and the breadth of victim geographies and business sectors. Table~\ref{tab:vendor_characterization} reports counts of vendors grouped by total number of reports in \datasetname{} and by the number of distinct TAs, geographies, and sectors they cover. We observe that {84.9\% (1,636) of}
vendors fall below 10 in all three dimensions, and {22.0\% (423)} of the vendors do not track any TA, geography, or sector. Only a small subset of {31} vendors publish more than 100 reports. Among these, {17} cover more than 100 TAs and {6} vendors exceed 100 reports, geographies and tracked TAs.
These figures suggest a highly fragmented CTI vendor landscape in which a reduced number of entities cover a large number of geographies and TAs, while {the majority track fewer than 10 geographies, TAs, and sectors.}

\begin{table}[t]
\caption{Percentage (and count) of vendors having different numbers of reports, threat actors, geographies and sectors.}
\label{tab:vendor_characterization}
\centering
\resizebox{\linewidth}{!}{%
\begin{tabular}{lrrrrr}
\toprule
\# Vendors & 0 & 1 & 2-9 & 10-99 & $\geq$100 \\
\midrule
\# reports & N/A & 54.8\% (1,055) & 33.9\% (653) & 9.7\% (187) & 1.6\% (31) \\
\# Threat actors & 38.0\% (731) & 18.5\% (357) & 32.3\% (623) & 10.2\% (196) & 1.0\% (19) \\
\# Geographies & 50.8\% (979) & 13.0\% (250) & 26.7\% (515) & 9.1\% (176) & 0.3\% (6) \\
\# Sectors & 27.8\% (536) & 24.6\% (474) & 40.3\% (776) & 7.3\% (140) & 0.0\% (0) \\
\bottomrule
\end{tabular}
}
\end{table}

\vspace{1mm}
\noindent\textbf{Evolution.}
We now study whether vendor coverage remains homogeneous over time or exhibits significant disparities, providing insights into how the industry's focus and visibility capabilities have evolved in response to a shifting threat landscape. Figure~\ref{fig:vendor_year_TA_reports_20} illustrates the activity of the top 25 vendors within \datasetname.
The data reveals a clear intensification in CTI dissemination, particularly within the last decade (as discussed in \ref{section:ecosystem:reporting}), alongside significant strategic divergence among vendors. Some vendors, such as $V_{1}$, $V_{2}$ or {$V_{4}$}, exhibit a strong correlation between high publication cadence and broad actor tracking (notice the large and warm-toned bubbles). In contrast, the presence of large yet cooler-toned nodes indicates broad detection capabilities distributed across fewer consolidated vendors, like {$V_6$, $V_9$ or $V_{13}$}. These results may be a consequence of the varying telemetry and reporting policies employed within the ecosystem. Conversely, there are cases like {$V_{5}$, $V_{8}$ or $V_{12}$}, with a large number of reports per year, but covering a reduced number of TAs in those reports.

\begin{figure}[t]
\centering
\includegraphics[width=0.9\linewidth]{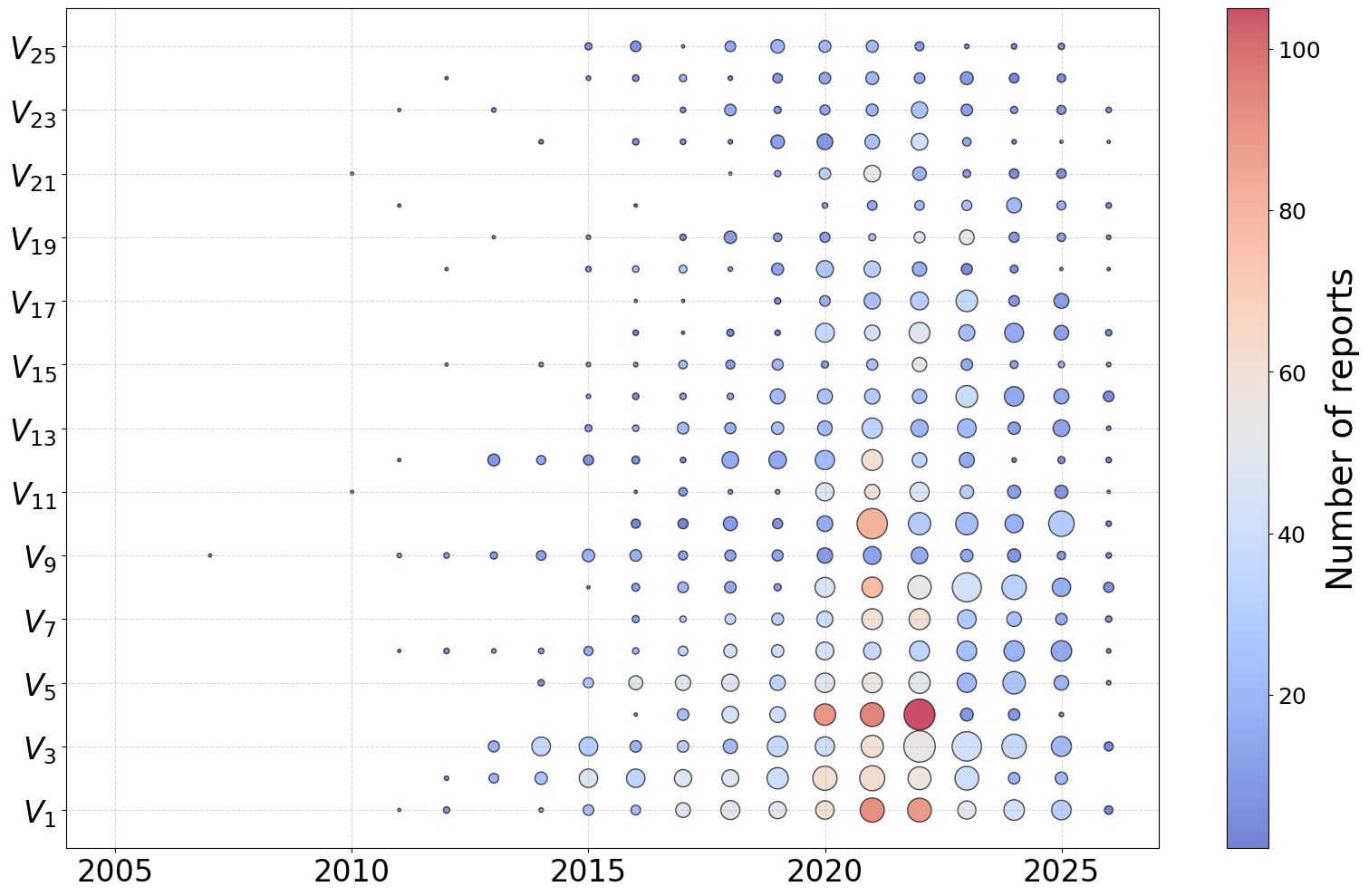}
\caption{Temporal evolution of the number of tracked actors (size of the bubble) and number of reports (color of the bubble) for the top 25 vendors.}
\label{fig:vendor_year_TA_reports_20}
\end{figure}

\subsection{Vendor Specialization}
\label{section:vendor:specialization}

The findings discussed in \S\ref{section:vendor:activity} raise questions about the relationship between vendors and the set of threat actors and geographies they track.
Vendor specialization is plausible because CTI visibility is inherently constrained by each vendor's sources and regional footprint. Therefore, analyzing the homogeneity of actor tracking across vendors is relevant to elicit the presence of observation biases.

\begin{figure}[t]
\centering
\includegraphics[width=0.95\linewidth]{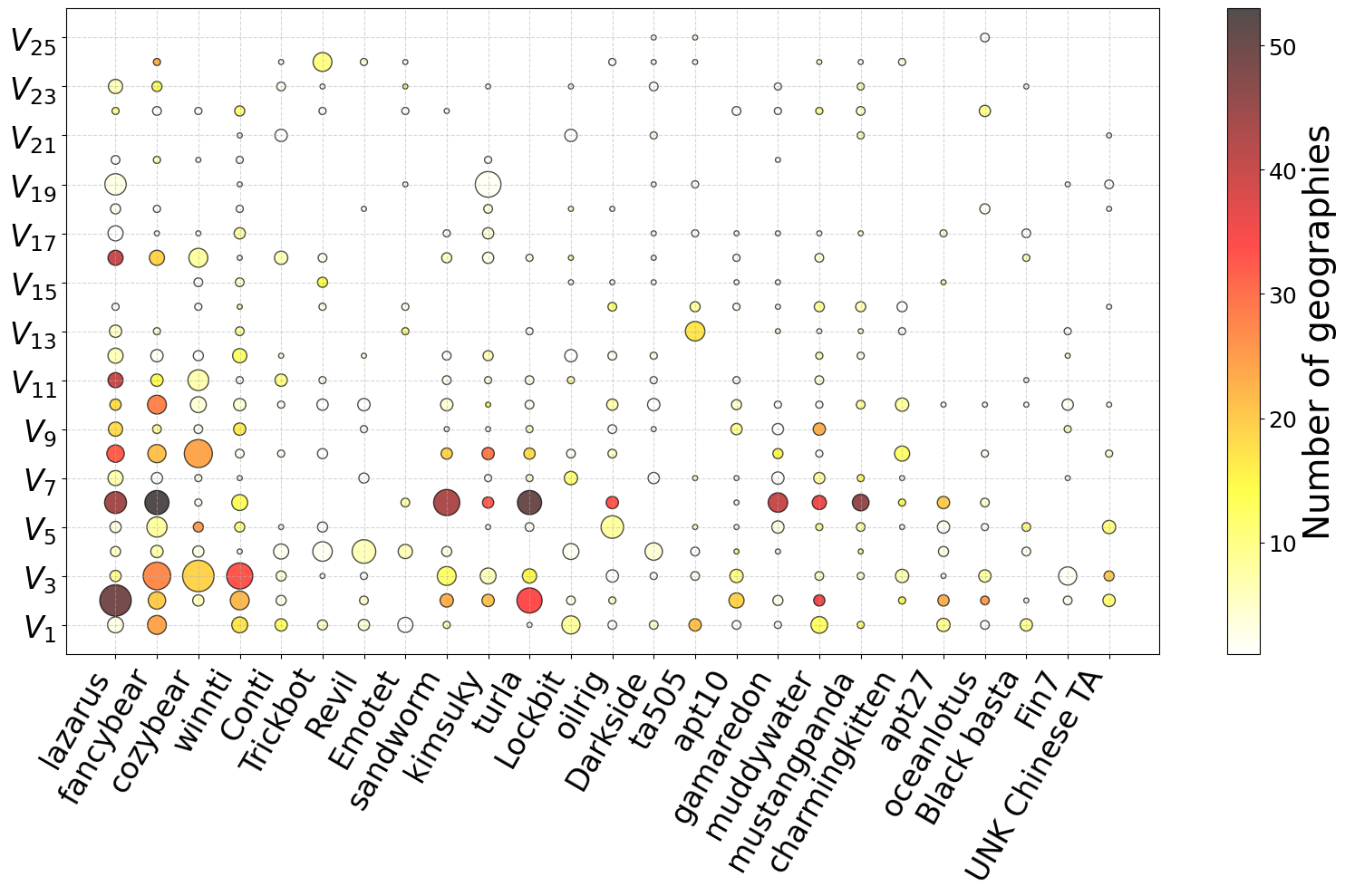}
\caption{Vendor-actor tracking relationship for the top 25 vendors and actors. The size of the bubble represents the number of reports about the actor and color represents the number of different geographies described.}
\label{fig:vendor_TA_reports_country}
\end{figure}

Figure~\ref{fig:vendor_TA_reports_country} shows the activity of the top 25 CTI vendors related to the top 25 actors.
We observe a noticeable reporting asymmetry, with visibility into major adversaries like \texttt{Lazarus} or \texttt{Fancy Bear} not being uniform across the industry. While some vendors like {$V_{3}$ and $V_{6}$} maintain broad coverage across
geographies, others exhibit highly localized specialization for certain actors: see, for example, $V_{19}$'s high-volume reporting activity on \texttt{Lazarus} and \texttt{Kimsuky}; $V_{13}$'s intensive focus on \texttt{TA505}; or $V_{24}$'s concentration on {\texttt{Trickbot}}. 
Looking beyond specific pairs of vendors and actors, we find that the full vendor-actor-geography-sector distribution is largely sparse. This confirms that not every vendor tracks the same threat actors, nor even the same victims of those actors, and therefore no single vendor provides a comprehensive view of the threat actor landscape.

\vspace{1mm}
\noindent\textbf{Marginal coverage analysis.}
Given the findings on vendor specialization, we perform a marginal utility analysis to determine how many vendors are required to achieve comprehensive coverage of known actors or victim geographies and sectors.
Figure~\ref{fig:vendor_ta_cobertura_all} shows the cumulative distribution of threat actors and geography-sector pairs covered by vendors, ranked in descending order by their total report volume. This distribution captures the marginal utility of integrating additional vendors into a CTI portfolio.
There is a noticeable divergence in coverage dynamics between the two metrics. The sector-geography dimension exhibits a quicker saturation. The top 10 vendors alone are able to capture approximately {77.8}\% of the landscape, and coverage reaches 90\% with the inclusion of the top 50 vendors. In contrast, threat actor coverage grows at a significantly slower rate. The top 10 vendors account for only {43.0}\% of the actor space, and even with 50 vendors it remains below {70}\%. This difference suggests that while a small core of major vendors provides broad geographic and sectoral visibility, effectively tracking the long tail of diverse threat actors requires an extensive vendor ecosystem.

\begin{figure}[t]
\centering
\includegraphics[width=0.95\linewidth]{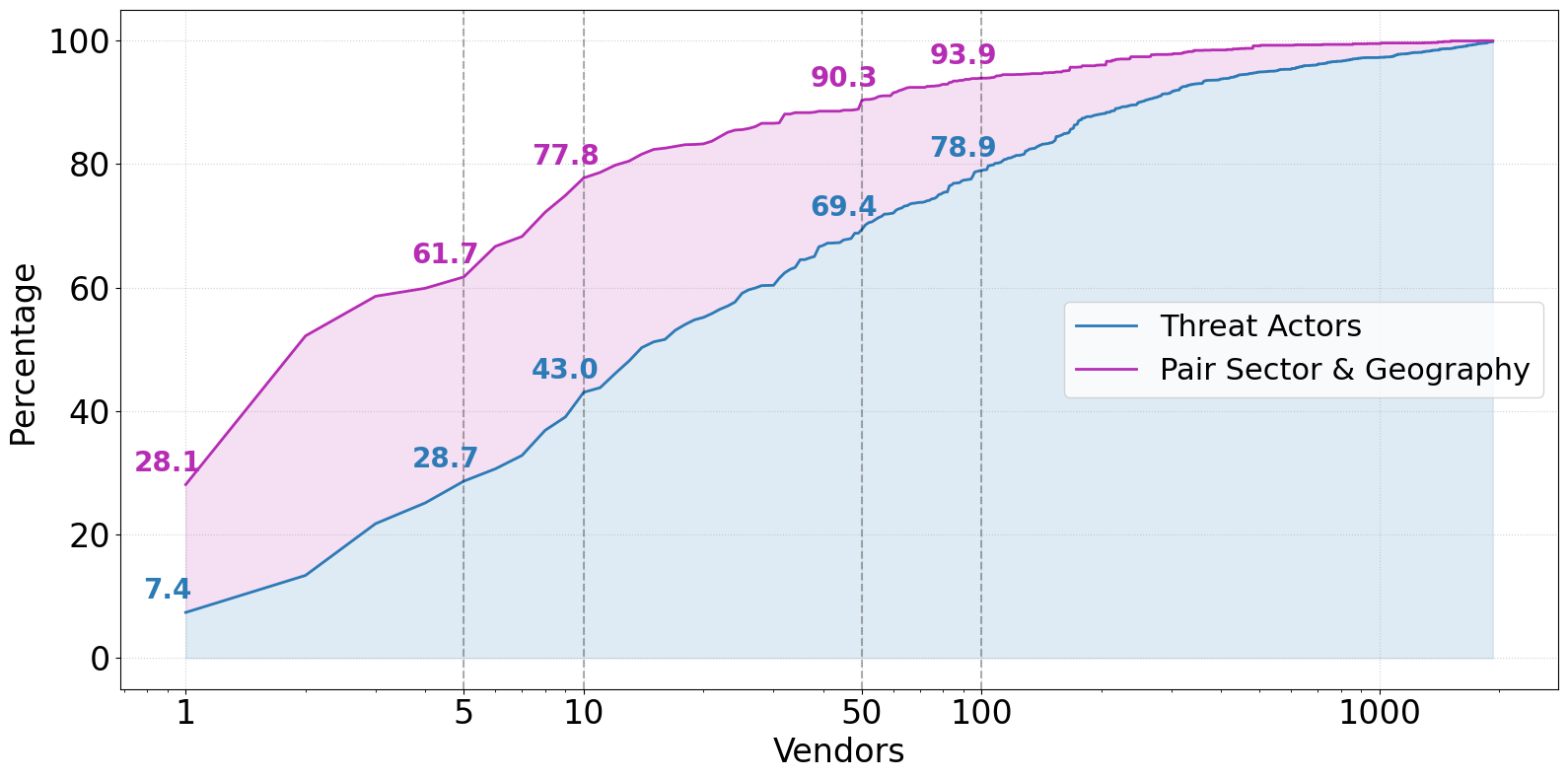}
\caption{Cumulative coverage curve for threat actors and geography-sector pairs.}
\label{fig:vendor_ta_cobertura_all}
\end{figure}

\subsection{Vendor Intelligence Overlap}
\label{section:vendor:overlap}

This section studies a complementary aspect to the findings about vendor specialization reported in \S\ref{section:vendor:specialization}, namely the degree of intelligence redundancy across them.

If two vendors provide extensive intelligence on a TA but their sources exhibit high overlap, the marginal utility of integrating both sources may be negligible. Conversely, a vendor with a smaller footprint that offers unique, non-overlapping insights becomes a valuable asset for achieving broader coverage. To study intelligence overlap, we consider data points through four strategic dimensions: geographies, sectors, motivations, and TTPs. In a separate analysis, we add IoCs to study vendor overlap when including tactical artifacts.

\begin{figure}[t]
    \centering
    \includegraphics[width=0.8\linewidth]{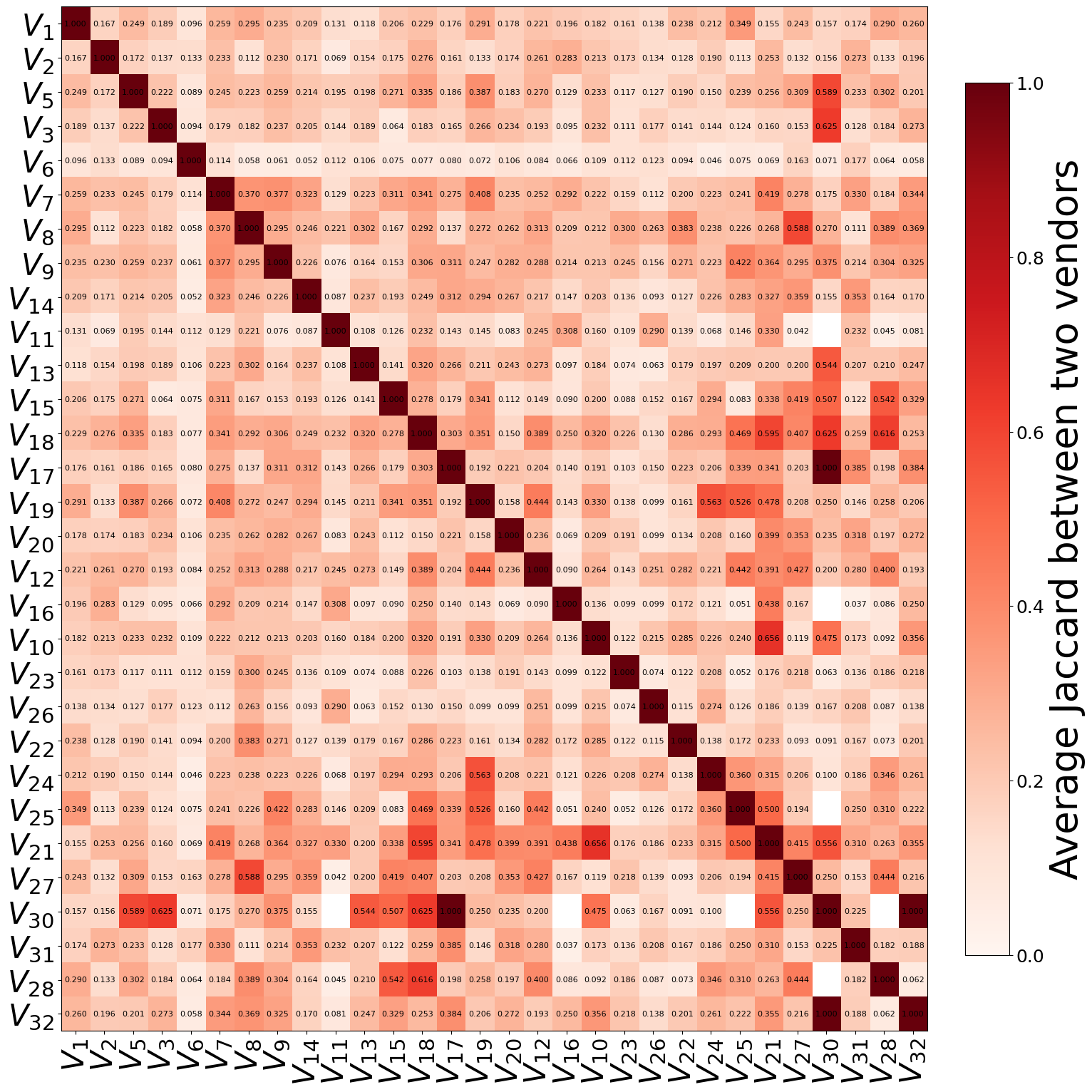}
    \caption{Jaccard of common TAs tracked among top 30 vendors with more reports in \subsetname.}
    \label{fig:heatmap_Jaccard_average}
\end{figure}

Figure~\ref{fig:heatmap_Jaccard_average} shows the intersection of intelligence data points among the top 30 vendors in \subsetname{} through the average Jaccard index for shared TAs. The results reveal a heterogeneous landscape where most vendor pairs exhibit relatively low overlap. This suggests that even dominant players maintain distinct observational vantage points. Vendors such as {$V_{21}$ (33.6\%) and $V_{18}$ (32.1\%)
show the highest overlap with the rest of the top 30.
Interestingly, they also show the highest overlap across all vendors (all of them, not just the top 30), but the overlap is much lower on average: 14.9\% and 13.5\%, respectively. In contrast, the third one, $V_{30}$, is ranked 76th (5.6\%) among vendors with the highest average overlap between each vendor.
This finding is not particular to these three vendors: the average overlap among vendors in the top 30 is {22.2}\%, while the average overlap between vendors in \subsetname{} is {1.5}\%.
Additionally, certain pairs, such as {$V_{30}$-$V_{17}$, $V_{21}$-$V_{10}$ or $V_{18}$-$V_{28}$}, with overlaps above 60\%, or even clusters like {$V_{19}$, $V_{21}$ and $V_{25}$} with Jaccard indexes {over 47}\%, show high similarity, likely due to a shared visibility into threats. This observation emphasizes the importance of identifying vendors with non-redundant telemetry to drastically improve an actor profile.
To quantify this divergence, we compute the average Jaccard similarity across the top $N$ (Figure~\ref{fig:overlap_N_vendors}). Even when restricting the analysis to vendor pairs with shared actors, similarity remains low, peaking at {0.28} for intelligence data points and falling below 0.1 when adding IoCs. These results are in consonance with those recently reported in \cite{AptToDisagree2026Eeten} using a private dataset of IoCs. As $N$ increases, the global average similarity tends to zero despite the rise in absolute actors intersections. This complementarity supports the need for multi-vendor fusion for wide visibility.

\begin{figure}[t]
    \centering
    \includegraphics[width=\linewidth]{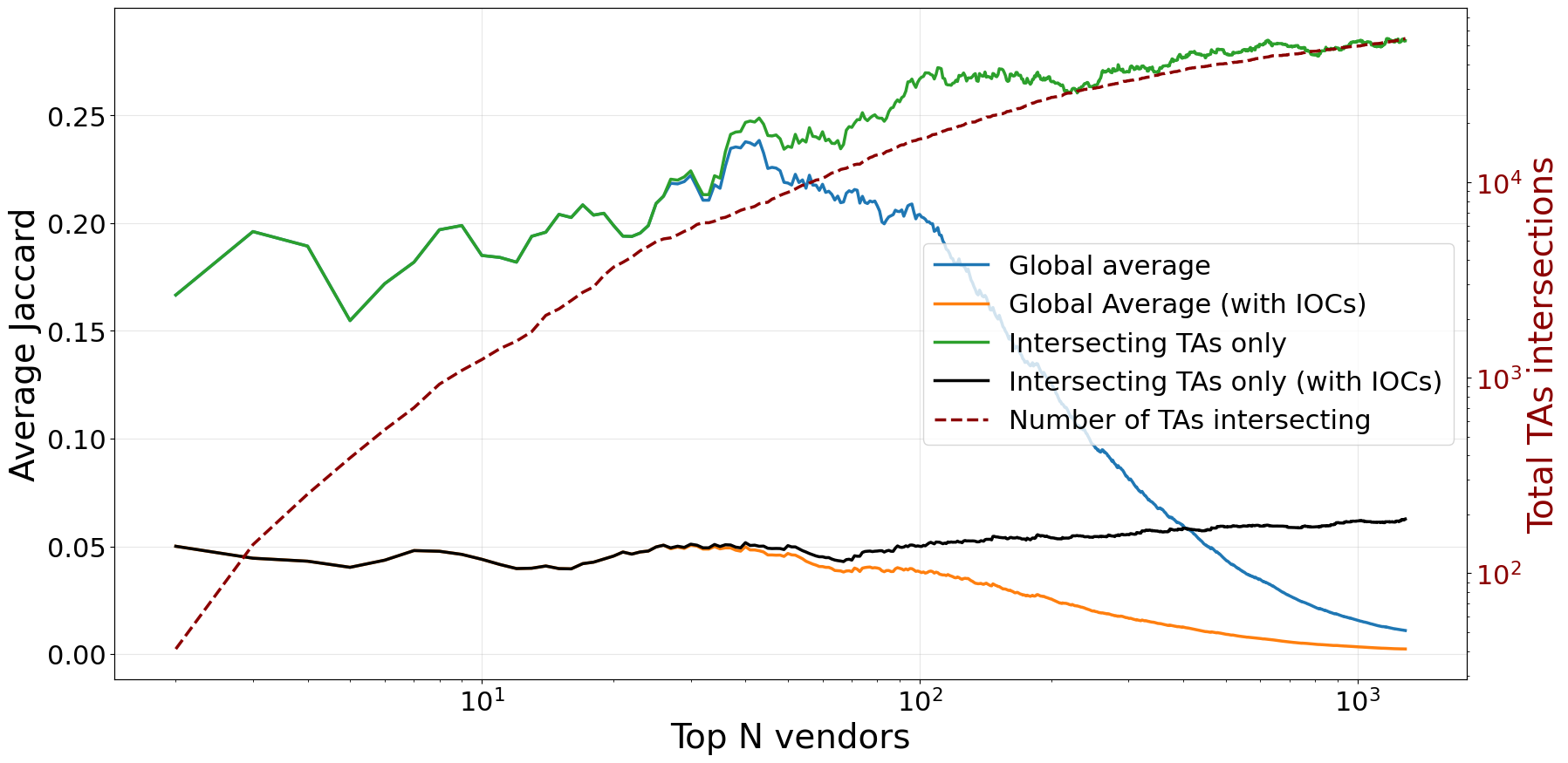}
    \caption{Average Jaccard index among the top $N$ vendors in \subsetname{} and number of cumulative common threat actors among those $N$ vendors.}
    \label{fig:overlap_N_vendors}
\end{figure}

\vspace{1mm}
\noindent\textbf{Marginal intelligence gain analysis.}
To quantify more accurately the marginal utility of vendor aggregation, we study the completeness of threat visibility across the 100 most frequent actors in \subsetname{} as a function of vendor aggregation.
We obtain for each actor the percentage of unique intelligence data points covered by the optimal subset of $N$ vendors (specifically, the $N$ vendors contributing the highest volume of non-redundant indicators relative to the total pool of available data).
Figure~\ref{fig:coverage_vendors_TA} shows the results for different numbers of vendors, restricting the number of actors to the top 100. The superimposed boxplots show the statistical distribution, allowing us to study the variance in vendor visibility for different threat actors. The color of each point in the scatter plot denotes the position of that TA in the ranking of most frequent actors.
The data reveals a high degree of intelligence fragmentation within the CTI ecosystem. Reliance on a single source ($N=1$) yields a median coverage below {25}\%, with significant heterogeneity across TAs, suggesting that individual vendors possess strictly partial views of the threat landscape. This phenomenon is amplified for the most frequent TAs, which exhibit an even lower coverage. The logarithmic growth of the curve indicates diminishing marginal utility of aggregating vendors. Notably, the system only approaches 100\% coverage when aggregating approximately {15} distinct sources. Combined with previous findings, this evidence suggests that achieving comprehensive situational awareness requires the integration of data from multiple vendors. Because the vendor subset is selected optimally for each actor, these percentages should be interpreted as an upper bound.

\begin{figure}[t]
\centering
\includegraphics[width=\linewidth]{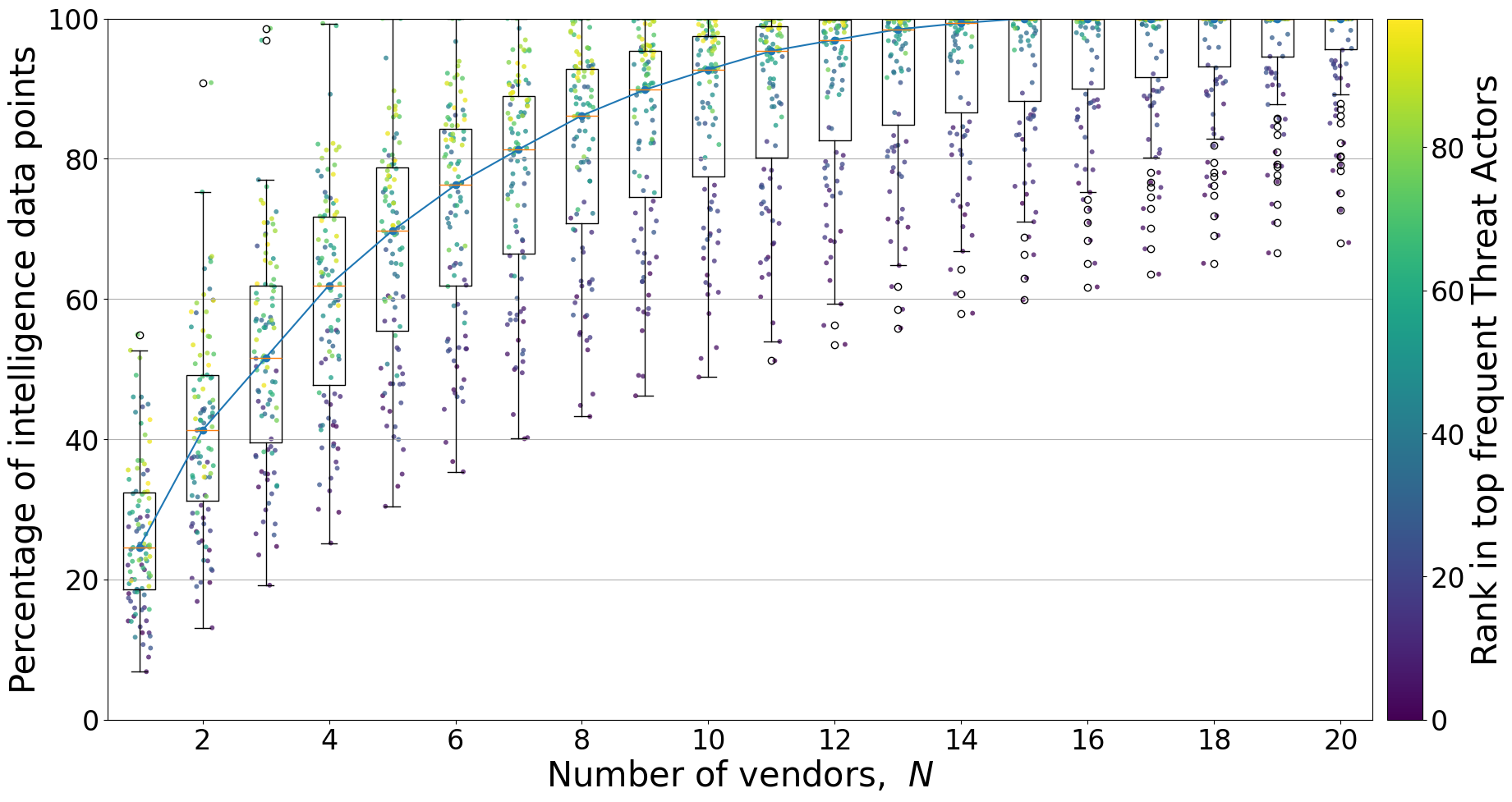}
\caption{Cumulative coverage of unique intelligence data points for the top 100 threat actors. 
}
\label{fig:coverage_vendors_TA}
\end{figure}

\begin{graybox}
\textbf{Takeaway.}
The CTI market is a highly fragmented, long-tail ecosystem where {85}\% of vendors are niche players and a small set of super-vendors provide the bulk of global, multi-actor intelligence. While IoC reporting is the industry baseline, TTPs are reported significantly less often and by very few vendors.
The average intelligence overlap between any two vendors is very low, both regarding the threat actors they track and the specific intelligence provided about a common actor. This suggests that a few core providers may suffice for broad situational awareness, but deep intelligence (especially on specific adversaries) may require a highly diversified multi-vendor strategy.

\end{graybox}

\section{Threats to Validity}
\label{sec:limitations}

\vspace{1mm}
\noindent\textbf{Reporting biases.}
The sources used to build \datasetname{} are exclusively open-source, primarily commercial, and largely based in Western countries. The reports that we analyze are written with vendor-specific goals and may require substantial human effort, which incentivizes vendors to strategically publish findings that differentiate themselves. These factors introduce biases regarding the information (un)available in the dataset and may influence the insights derived from our analysis. %
We acknowledge that such biases are inherent to open-source CTI research.
Additionally, our visibility is limited to public CTI, and therefore we do not study private intelligence. Nevertheless, open-source reports remain the primary data source for recent academic studies of CTI~\cite{Yuldoshkhujaev2025,saha2026kittenpandameasuringspecificity, Buchel25}.

\vspace{1mm}
\noindent\textbf{Coverage.}
While \subsetname{} is extensive, its coverage of the complete open-source CTI landscape is unknown due to the lack of ground truth. In an effort to estimate our coverage, we adopt the methodology proposed in \cite{saha2026kittenpandameasuringspecificity}, focusing on the top vendors by report volume during 2021. We manually audit their publications and separate CTI reports from other content found on their websites.
The results (Table~\ref{tab:vendor_coverage}) show an average coverage of 57.7\%, with vendor-specific rates ranging from 44.3\% to 73.0\%. Although these figures reflect the challenges of exhaustive CTI collection, they suggest that \subsetname{} captures a representative and significant part of the ecosystem.

\begin{table}[t]
\caption{Coverage analysis across top CTI vendors.
}
\label{tab:vendor_coverage}
\centering
\resizebox{0.9\linewidth}{!}{%
\begin{tabular}{lrrrr}
\hline
Vendor & \# Entries & \# Reports & In \subsetname & \% \\
\hline
$V_{1}$ & 128 & 75 & 44 & 58.7\%\\
$V_{2}$ & 70 & 37 & 27 & 73.0\% \\
$V_{5}$ & 100 & 69 & 37 & 53.6\% \\ 
$V_{3}$ & 56 & 39 & 26 & 66.7\% \\
$V_{6}$ & 273 & 61 & 27 & 44.3\% \\
$V_{7}$ & 86 & 48 & 27 & 56.3\% \\
$V_{8}$ & 156 & 47 & 29 & 61.7\% \\
\hline
All & 869 & 376 & 217  & 57.7\%\\
\hline
\end{tabular}
}
\end{table}

\vspace{1mm}
\noindent\textbf{IoC extraction.}
IoCs are the most technically dense information type contained in \datasetname{} (\S\ref{section:ecosystem:reporting}). We deliberately rely exclusively on LLMs rather than dedicated IoC extraction utilities~\cite{iocparser_paloalto, inquest_iocextract, Caballero2023}, 
despite such tools being widely adopted by the community~\cite{Furumoto2025, Yuldoshkhujaev2025, Kumarasinghe2024}.
Predominantly regex-based approaches
frequently over-extract artifacts (most notably URLs) by automatically labeling links embedded in reports as IoCs, which constitutes a clear limitation.
To reduce false positives, our approach leverages contextual inference to distinguish report metadata
from actionable IoCs.

\vspace{1mm}
\noindent\textbf{Normalization.}
Our pipeline includes manual review to construct taxonomies and normalization mappings, as well as for duplicate items (see \S\ref{sec:labeling}, \S\ref{sec:mappings}, and \S\ref{sec:methodology:duprem}). We acknowledge that this manual processing limits scalability. However, as discussed in \S\ref{sec:relatedwork}, this step is essential to mitigate the substantial level of noise in the extracted data if LLMs are given total freedom to populate data fields. Our experiments without these steps resulted in data with significantly less analytic value because of the diversity of values.
Future work should focus on standardizing and automating this phase. In the case of TA names, alternative alias mappings could affect our observations, in particular the marginal coverage analysis. Unfortunately, TA naming remains an open problem and many TA names 
are not traceable to any vendor or source~\cite{roa2025hesperusphosphorusmappingthreat}.

\vspace{1mm}
\noindent\textbf{CTI scope.}
This work focuses on long-form technical blogs and reports. We acknowledge that substantial CTI also appears in other open sources, including social media posts, specialized feeds, 
and aggregators/platforms (e.g., VirusTotal), structured and semi-structured formats such as STIX, and GitHub repositories. We leave their integration to future work.

\section{Conclusions}
We presented the creation and analysis of a large dataset of 16,096 CTI reports published over the last 20 years. We believe this is the largest publicly released collection analyzed to date. Our methodology uses LLMs along with safeguards to ensure that the features we extract are accurate. The analysis of the dataset provides insights into the evolution of the reported threat landscape and the CTI reporting ecosystem, the connections between threat actors and victims, and the observational biases caused by the vendor effect. Our findings suggest that researchers and practitioners should treat public CTI not as a neutral sample of the threat landscape, but as a vendor-mediated measurement layer with measurable blind spots.

\section*{Ethical Considerations}
Our data collection methodology was guided by the following ethical principles to ensure responsible research conduct.

\vspace{1mm}
\noindent\textbf{Infrastructure respect.}
We performed all web scraping of CTI report files with diligence to ensure negligible impact on source infrastructure. By implementing rate-limiting and spacing out requests, we strictly adhered to standard crawling practices, respecting server workloads and avoiding any disruption of service for the hosting entities.

\vspace{1mm}
\noindent\textbf{Privacy and Personally Identifiable Information (PII).}
The scope of our collection was limited to publicly available, open-source CTI reports. The dataset we publish does not contain Personally Identifiable Information (PII) and focuses exclusively on technical indicators about cyberattacks, threat actor behaviors, and CTI vendors.

\vspace{1mm}
\noindent\textbf{Legal aspects.}
To respect intellectual property and copyright constraints, we do not redistribute the original source documents. Instead, we provide only the derived annotated metadata. This approach ensures the reproducibility of our findings while honoring the proprietary rights of the original publishers.

\bibliographystyle{IEEEtran}
\bibliography{biblio}

\appendices

\section{Open Science}
\label{sec:appendix:openscience}
We make publicly available all relevant research artifacts in the following repository:

\vspace{2mm}

\vspace{2mm}
\noindent The repository contains the following items:

\vspace{1mm}
\noindent\textbf{Prompts and taxonomies.}
We include the prompts used to classify the reports and extract the initial taxonomy (\S\ref{sec:labeling}) and to label all reports in the initial collection (\S\ref{sec:processing}). The prompts include two roles, one as an analyst and another as a developer for increased accuracy. The second prompt includes the taxonomy used for designated fields. We also include the JSON schemas describing the taxonomies for the \texttt{Enum} fields used to label all reports in \datasetname. 
We do not explicitly write them here due to their size.

\vspace{1mm}
\noindent\textbf{Normalization mappings.}
We publish the mappings for the threat actors, vendors, and geographies. These mappings are used in the normalization stages of our methodology (\S\ref{sec:mappings}). We also include the vendor taxonomies and mappings used to deduplicate the space of threat actor aliases.

\vspace{1mm}
\noindent\textbf{Validation artifacts.}
We include the output (JSON documents) produced by each of the models evaluated in \S\ref{sec:appendix_val:othermodels}, which are used to compare them to our base model.

\vspace{1mm}
\noindent\textbf{Datasets.}
We make available the complete \datasetname{} and \subsetname{} datasets. Each item in the dataset is a JSON document containing the extracted labels, along with other relevant metadata and links to the original reports (PDF files).

\section{Generative AI Usage}
 Generative AI tools constitute a core component in the construction of \datasetname{}, as they are used to extract data from each report. The exact methodology is described in \S \ref{sec:methodology}.
 
 We also use AI tools to revise the text and perform grammatical checks. Each sentence was subsequently reviewed manually and revised as needed to more accurately reflect the authors' intended meaning.

\section{Dataset description}
\label{sec:appendix:dataset_description}
Table \ref{tab:dataset_description} shows the CTI fields extracted from each report in \datasetname~ and their data types.

\section{Processing cost}
\label{sec:appendix:cost}
We process the entire collection of reports using OpenAI's batch API~\cite{openaiBatchAPIPlatform}, which allows us to ``send asynchronous groups of requests with 50\% lower costs, a separate pool of significantly higher rate limits, and a clear 24-hour turnaround time''~\cite{openaiBatchAPIPlatform}. Each of the batches contains 500 requests. We only extract the metadata from one report for each request to prevent responses from multiple reports from being mixed up or confused. OpenAI charges per token and both input and output tokens determine the processing cost. Input and output tokens have different costs and pricing also varies by model. For \datasetname, an average extraction request for each report consumes 62,493 input tokens and 3,552 output tokens. We do not observe a strong correlation between the word count of a report or its file size and the total number of tokens. Overall, 25\% of reports incur fewer than 27k tokens, whereas the most expensive quartile exceeds 71k tokens.

\section{Taxonomies}
\label{sec:appendix_taxon}
Table \ref{table:taxonomies} shows the two-level taxonomies built for the type of report, attack motivation, and attacked sectors.

\section{Validation}
\label{sec:appendix_val}
This appendix provides additional details regarding the validation of the methodology used to construct \datasetname.

\subsection{Precision, Recall, and F1-scores}
\label{sec:appendix_val:precrec}
Tables~\ref{tab:precision_and_recall_selected_reports} and ~\ref{tab:f1_selected_reports} provide the precision, recall and F1-scores obtained during the validation of our methodology (see \S\ref{sec:validation}).

\subsection{Comparison with Other LLMs}
\label{sec:appendix_val:othermodels}

To further evaluate the robustness and reproducibility of our methodology (\S\ref{sec:methodology}), we benchmark \modelname{} \texttt{(o3)} against several state-of-the-art  LLMs with varying architectures and parameter scales. Our selection criteria prioritize models frequently used in recent CTI research. The GPT-4 family is a dominant baseline~\cite{Mezzi2025, Bhusal2024, Tanvirul24, Tihanyi2024, Schwartz2025, Cheng2025, Yang2026, wang2024multikgmultisourcethreatintelligence, Yuldoshkhujaev2025}, leading us to select \texttt{gpt-4o (4o)} (snapshot \texttt{2024-11-20}) as it is the current top performer in this family~\cite{openaiHolaGPT4o}. 
To show cross-provider diversity, we also evaluated Google's Gemini catalog, incorporating \texttt{Gemini 2.5 (G2.5F)} (the successor to the now-deprecated \texttt{Gemini 1.5} that is frequently used in recent works~\cite{Mezzi2025, Yuldoshkhujaev2025}) and \texttt{Gemini-3-pro-preview (G3PP)}, the current leader of the LMSYS leaderboard~\cite{lmarena_text_arena_2026_01_29}. We further include \texttt{Gemma 3 (27B) (G3)} to represent high-performance open-weight alternatives. Each model was evaluated using the same set of 100 validation reports to ensure experimental parity.

As illustrated in Table~\ref{tab:validation_results}, \modelname{} consistently outperforms both proprietary and open-weight baselines. While \texttt{Gemini-3-pro} exhibits slightly superior performance in specific fields, \modelname{} remains remarkably competitive despite the novelty of this model. We attribute this to current limitations in Google’s API (like structured output enforcement) compared to OpenAI's ecosystem. This reinforces our hypothesis that frontier LLM reasoning must be augmented by rigorous pre- and post-processing pipelines as detailed in \S\ref{sec:methodology} to ensure reliability.

\section{Top Targeted Countries, Business Sectors, and Motivations}
\label{sec:appendix:topvictims}
Table~\ref{tab:top_countries} provides key statistics about the top countries or geographies targeted by cyberattacks disaggregated by attack motivation and business sector.

\begin{table}[t]
\caption{Labels extracted from each report in \datasetname. See \S\ref{sec:labeling} for a full description and details about the enum types.}
\label{tab:dataset_description}
\centering
\begin{tabular}{ll}
\toprule
Field & Type\\
\midrule
Title & String\\
Report Year & Timestamp\\
Report Month & Timestamp\\
CTI Vendor & String\\
Report Type & Enum\\
Report Subtype & Enum\\
Attack Motivation & Enum\\
Attack Submotivation & Enum\\
Attacked Business Sector & Enum \\
Attacked Business Subsector & Enum \\
Victim & String \\
Attacked Country/Geography & String\\
Threat Actor & String\\
Campaign Duration & Time range\\
Indicators of Compromise (IoCs) & List\\
Tactics, Techniques, and Procedures (TTPs) & List\\
\bottomrule
\end{tabular}
\end{table}

\begin{table}[ht!]
\caption{Validation scores for each LLM model and our choice (\texttt{o3}). Best score for each field is shown in \textbf{bold}.}
\label{tab:validation_results}
\begin{tabular}{llllll}
\toprule
Field  & \textbf{\texttt{o3}} & \texttt{4o} & \texttt{G3PP} & \texttt{G2.5F} & \texttt{G3}\\ 
\midrule
Title                      & 95.0  & 75.9 & \textbf{98.8} & 95.7 & 97.2 \\
Report Year                & \textbf{97.2}  & 90.9 & 97.0 & 95.2 & 93.2\\
Report Month               & \textbf{98.1}  & 83.5 & 94.6 & 91.6 & 91.0 \\
CTI Vendor                 & 98.1  & 90.8 & \textbf{99.4} & \textbf{99.4} & 92.7\\
Report Type                & \textbf{94.6}  & 61.9 & 86.9 & 62.7 & 72.3  \\
Report Subtype             & \textbf{94.1}  & 56.8 & 85.1 & 57.8 & 68.1 \\
Motivation                 & \textbf{87.8}  & 71.5 & 86.8 & 85.9 & 79.8  \\
Submotivation              & \textbf{86.6}  & 68.9 & 83.4 & 68.9 & 68.5  \\
Sector                     & \textbf{91.1}  & 60.6 & 91.1 & 50.5 & 50.8  \\
Subsector                  & \textbf{92.4}  & 56.3 & 90.0 & 47.8 & 40.3  \\
Victim                     & \textbf{96.6}  & 89.1  & 91.4  & 85.9   & 84.0 \\
Geography                  & \textbf{94.3}  & 82.5 & 88.7 & 93.6 & 84.0\\
Threat Actor               &  96.0 & 83.9 & \textbf{96.6}  & 85.3 & 87.9 \\
Campaign Duration            & \textbf{92.4}  & 38.3 & 80.0  & 33.3 & 59.6 \\
IoCs                       & 88.6  & 56.1 & \textbf{91.7} & 86.5 & 73.1 \\
ATT\&CK TTPs               & \textbf{98.1}  & 49.1 & 83.1 & 86.4 & 76.2  \\
\midrule
Overall          & \textbf{93.8} & 69.8  & 90.3  & 76.7 & 76.2     \\           
\bottomrule
\end{tabular}

\end{table}

\newcolumntype{L}{>{\raggedright\arraybackslash}X}

\begin{table*}[th!]
\centering
\small
\caption{Taxonomies for the report types, target sectors, and motivations.}
\label{table:taxonomies}
\begin{tabularx}{\textwidth}{@{} c l l L @{} }
\toprule
Tax. & Acronym & Level 1 (Category) & Level 2 (Subcategories) \\ \midrule

\multirow{12}{*}{\rotatebox{90}{Motivations}} 
& AE & Accidental / Error & Software error, User or Administrator Error \\
& EIC &Espionage \& Intelligence Collection & Espionage, Intellectual Property Theft \\
& FC &Financial / Cybercrime & Blackmail, Credential Theft, Data Theft, Financial Fraud, Financial Gain \\
& ICP & Illicit Content \& Perversion & Child Pornography, Perversion, Piracy \\
& MCW & Military \& Cyber Warfare & Military, Cyber Warfare \\
& MP & Malware Propagation & Malware Infection \\
& OTH & Other & Other \\
& PER & Personal & Personal \\
& PII & Political \& Ideological Influence & Hacktivism, Political, Propaganda \\
& RST & Reconnaissance \& Security Testing & Audit or pen test, Reconnaissance, Curiosity \\
& SAB & Sabotage & Sabotage \\
& SO & Strategic Objectives & Strategic \\ \midrule

\multirow{10}{*}{\rotatebox{90}{Report Types}} 
& ACA & Academic & Case Study, Whitepaper \\
& CA & Campaign Analysis & Campaign Analysis Report, Campaign Disruption Report \\
& CPL & Compliance, Policy \& Legal & Affidavit, Compliance \& Risks, Policy Analysis Report \\
& IHF & Incident Handling \& Forensics & Digital Forensic Analysis, Forensic Triage Methodology, Incident Report, Incident Response Report, Takedown Operation Report \\
& ITI & IoCs, TTPs \& Indicators & IoC Report, MITRE ATT\&CK Mapping \\
& MVA & Malware \& Vulnerability Analysis & Malware Analysis Report, Malware Family Profile, Exploit Analysis, Vulnerability Report, Vulnerability Advisory \\
& OMC & Outreach, Media \& Communications & Tool Documentation, Threat Hunting Tutorial, News Article, Press Release, Tutorial \\
& OTH & Other & Other \\ 
& TAA & Threat Actor Analysis & Threat Actor Activity Analysis, Threat Actor Attribution, Threat Actor Profile, Threat Actor Retirement Announcement \\
& TLT & Threat Landscape \& Trends & Periodic Threat Landscape Report, National Cybersecurity Organization Report, Threat Trend Report \\\midrule

\multirow{10}{*}{\rotatebox{90}{Sectors}} 
& AAS & Aerospace, Aviation \& Space & Aerial Imagery, Aerospace, Civil Aviation, Space \& Satellite Technology \\
& CF & Cryptocurrency \& Fintech & Cryptocurrency, Fintech, Darknet Marketplace \\
& CI-ICS & Critical Infrastructure \& ICS/OT & Critical Infrastructure, Industrial Control Systems, Operational Technology \\
& CPE & Consulting \& Professional Services & Business, Consulting, Conglomerates\\
& ER & Education \& Research & Higher education, Other Education, Scientific research \\
& ESG & Entertainment, Sports \& Gambling & Entertainment, Gambling, Gaming, Music Industry, Sports\\
& EUNR & Energy, Utilities \& Natural Resources & Basic Resources, Energy, Mining, Natural Resources, Nuclear Industry, Oil and Gas, Utilities, Water \\
& FA & Food \& Agriculture & Agriculture, Food Production, Food and Beverage\\
& FBI & Finance, Banking \& Insurance & Accounting, Asset Management, Finance, Insurance, Investment Banking, Online Banking \\
& GPA & Government \& Public Administration & Government Agencies, Municipal Government, Customs, Foreign Affairs, Diplomacy, Political, Prisons, Emergency Services \\
& HLS & Healthcare \& Life Sciences & Biomedical, Biotechnology and Pharmaceutical, Health, Veterinary Services \\
& HT & Hospitality \& Travel & Hotels, Travel\\
& IPS & Individuals \& Personal Services & Individuals, Job-seeking\\
& ITTC & IT, Telecom \& Cybersecurity & Cloud Service Provider, Cybersecurity, IT services, Physical Security, Email Providers, Internet Infrastructure, Social Media, Telecom \\
& LJ & Legal \& Judiciary & Judiciary, Legal\\
& MAJ & Media, Advertising \& Journalism & Advertising, Journalism, Media, Internet Media, Press Services\\
& MIL & Military & Military, Intelligence \\
& MIM & Manufacturing, Industrial \& Materials &  Automotive, Chemical, Electronics, Manufacturing, Materials, Printing, Semiconductor, Supply Chain\\
& NACS & NGOs, Advocacy \& Civil Society &  Advocacy, Civil Society, Human Rights, Non-governmental Organization, Non-profit, Think Tank\\
& OTH & Other & Other \\ 
& RC & Religion \& Culture & Cultural, Religious\\
& RCCS & Retail, Commerce \& Consumer Services & Consumer, E-commerce, Fashion, Point of Sale, Restaurants, Retail, Trade\\
& RECE & Real Estate, Construction \& Engineering & Construction, Engineering, Interior design, Real Estate\\
& TLM & Transportation, Logistics \& Maritime & Logistics, Maritime, Railroads, Shipping\\
\bottomrule
\end{tabularx}
\end{table*}

\begin{table*}[t]
\caption{Precision and recall (P/R) values obtained during validation disaggregated by field and type of report.}
\label{tab:precision_and_recall_selected_reports}
\centering
\resizebox{\textwidth}{!}{%
\begin{tabular}{rrrrrrrrrrrr}
\toprule
\multirow{2}{*}{Field} & \multicolumn{10}{c}{Type of report}             & \multirow{2}{*}{All} \\ \cmidrule{2-11}
& TLT  & TAA  & CA   & MVA  & IHF  & CPL & ACA   & ITI  & OMC  & Other &\\
& (5)  & (10) & (20) & (25) & (5)  & (5)  & (5)   & (5)    & (15)   & (5) & (100)\\ 
\midrule
Title                  & 100/100  & 100/100  & 96/96    & 100/100  & 100/100  & 100/100 & 80/80   & 90/80   & 80/90   & 75/75    & 94.8/95.3 \\
Report Year            & 100/100  & 100/100  & 100/100  & 100/100  & 100/100  & 100/100 & 100/100 & 60/60   & 100/100 & 75/75    & 97.2/97.2 \\
Report Month           & 100/100  & 92.9/92.9 & 100/100  & 100/100  & 100/100  & 100/100 & 100/100 & 100/100 & 100/100 & 75/75    & 98.1/98.1 \\
CTI Vendor             & 100/100  & 100/100  & 100/100  & 100/100  & 100/100  & 100/100 & 100/100 & 100/100 & 90/90   & 75/75    & 98.1/98.1 \\
Report Type            & 100/100  & 92.9/89.3 & 100/98   & 96.2/96.2 & 100/91.7 & 80/80  & 60/60   & 100/100 & 100/100 & 100/100   & 95.3/93.9 \\
Report Subtype         & 100/100  & 92.9/89.3 & 100/98   & 96.2/96.2 & 100/91.7 & 80/80  & 60/60   & 100/80  & 100/100 & 100/100   & 95.3/92.9 \\
Motivation             & 100/100  & 92.9/92.9 & 92/92    & 84.6/88.3 & 83.3/83.3 & 100/100 & 100/64 & 40/40   & 90/100  & 75/75    & 87.7/87.9 \\
Submotivation          & 83.3/100 & 85.7/82.1 & 92/90    & 84.6/86.4 & 83.3/83.3 & 100/100 & 100/84 & 40/60   & 90/100  & 75/75    & 85.8/87.4 \\
Victim Sector          & 83.3/77.8 & 100/100  & 100/96.7 & 92.3/96.2 & 100/100  & 80/80  & 60/48   & 100/100 & 70/70   & 100/100   & 91.5/90.8 \\
Victim Subsector       & 100/77.8 & 100/98.8 & 100/96.5 & 96.2/100 & 100/100  & 80/80  & 60/48   & 100/100 & 70/70   & 100/100   & 93.4/91.5 \\
Victim                 & 83.3/100 & 100/100  & 100/96   & 98.1/97.3 & 100/100  & 100/100 & 80/60 & 100/100 & 100/100 & 75/100   & 96.7/96.5 \\
Victim Geography       & 83.3/79.6 & 100/98.3 & 100/93.4 & 96.2/96.2 & 66.7/66.7 & 100/100 & 80/80 & 100/100 & 100/100 & 100/100   & 95.3/93.3 \\
Threat Actor           & 83.3/83.3 & 100/100  & 100/100  & 100/98.1 & 100/100  & 100/100 & 60/60 & 100/100 & 100/100 & 75/75    & 96.2/95.8 \\
Campaign Duration      & 83.3/83.3 & 92.9/85.7 & 92/92   & 100/96.2 & 100/100  & 100/100 & 60/60 & 80/80   & 100/100 & 100/100   & 93.4/91.5 \\
IoCs                   & 96.7/89  & 100/74.9 & 91.4/82.9 & 93.8/84.3 & 100/94.2 & 100/100 & 80/80 & 100/63.6 & 100/91 & 75/75    & 94.4/83.4 \\
ATT\&CK TTPs     & 100/100  & 92.9/92.9  & 96/96  & 100/100  & 100/100  & 100/100 & 100/100 & 100/100 & 100/100 & 100/100   & 100/100 \\
\midrule
Overall                & 93.5/93.2 & 96.4/93.6  & 97.7/95.7 & 96.1/96  & 95.8/94.4 & 95/95  & 80/74 & 88.1/85.2 & 93.1/94.4 & 85.9/87.5 & 94.6/93.4 \\
\bottomrule
\end{tabular}%
}
\end{table*}
\begin{table*}[ht!]
\centering
\caption{F1-scores and Shannon entropy obtained during validation, broken down by field and type of report.}
\label{tab:f1_selected_reports}
\begin{tabular}{rrrrrrrrrrrrl}
\toprule
\multirow{3}{*}{Field} & \multicolumn{10}{c}{Type of report}                                &  & \multirow{3}{*}{$H$} \\ \cmidrule{2-11}
                       & TLT  & TAA   & CA    & MVA   & IHF  & CPL & ACA    & ITI  & OMC   & Other & All &  \\
& (5)    & (10)   & (20)   & (25)   & (5)    & (5)   & (5)    & (5)    & (15)   & (5)     & (100)                  &                    \\ 
\midrule
Title                  & 100  & 100  & 96   & 100  & 100  & 100 & 80   & 84.7 & 84.7 & 75    & 95                   & 0                  \\
Report Year            & 100  & 100  & 100  & 100  & 100  & 100 & 100  & 60   & 100  & 75    & 97.2                 & 0                  \\
Report Month           & 100  & 92.9 & 100  & 100  & 100  & 100 & 100  & 100  & 100  & 75    & 98.1                 & 0                  \\
CTI Vendor             & 100  & 100  & 100  & 100  & 100  & 100 & 100  & 100  & 90   & 75    & 98.1                 & 0                  \\
Report Type            & 100  & 91.1 & 99   & 96.2 & 95.7 & 80  & 60   & 100  & 100  & 100   & 94.6                 & 0.029              \\
Report Subtype         & 100  & 91.1 & 99   & 96.2 & 95.7 & 80  & 60   & 88.9 & 100  & 100   & 94.1                 & 0.029              \\
Motivation             & 100  & 92.9 & 92   & 86.4 & 83.3 & 100 & 78   & 40   & 94.7 & 75    & 87.8                & 0.08               \\
Submotivation          & 90.9 & 83.9 & 91   & 85.5 & 83.3 & 100 & 91.3 & 48   & 94.7 & 75    & 86.6                 & 0.095              \\
Victim Sector          & 80.5 & 100  & 98.3 & 94.2 & 100  & 80  & 53.3 & 100  & 70   & 100   & 91.1                 & 0.086              \\
Victim Subsector       & 87.5 & 99.4 & 98.2 & 98.1 & 100  & 80  & 53.3 & 100  & 70   & 100   & 92.4                 & 0.086              \\
Victim                 & 90.9 & 100  & 98   & 97.7 & 100  & 100 & 68.6 & 100  & 100  & 85.7  & 96.6                 & 0                  \\
Victim Geography       & 81.4 & 99.1 & 96.6 & 96.2 & 66.7 & 100 & 80   & 100  & 100  & 100   & 94.3                 & 0.029              \\
Threat Actor           & 83.3 & 100  & 100  & 99   & 100  & 100 & 60   & 100  & 100  & 75    & 96                   & 0                  \\
Campaign Duration      & 83.3 & 89.2 & 92   & 98.1 & 100  & 100 & 60   & 80   & 100  & 100   & 92.4                 & 0.171              \\
IoCs                   & 92.7 & 85.6 & 86.9 & 88.8 & 97   & 100 & 80   & 77.8 & 95.3 & 75    & 88.6                 & 0                  \\
ATT\&CK TTPs     & 100  & 92.9  & 96  & 100  & 100  & 100 & 100  & 100  & 100  & 100   & 98.1                  & 0.166              \\
\midrule
Overall                & 93.3 & 95 & 96.5 & 96.0 & 95.1 & 95.0  & 76.9 & 86.6 & 93.7 & 86.7  & 93.8                   & 0.048\\             
\bottomrule
\end{tabular}%
\end{table*}

\begin{table*}[]
\caption{Top geographies by within-geography attack share, stratified by attacker motivation and victim industry sector. \textit{Total} counts the number of attacks with a particular motivation or sector; \textit{Geo} counts the number of distinct victim geographies with at least one attack for that motivation or sector. Percentages in parentheses denote, for each geography, the fraction of its attacks attributed to the specified motivation or sector. Geographies in \textbf{bold} are detected as outliers (Tukey's method). The rightmost columns report the average, the standard deviation and quartiles of per-geography percentage shares for each motivation and sector. Only geographies with more than 10 attacks were considered.}
\label{tab:top_countries}
\centering
\resizebox{\textwidth}{!}{
\begin{tabular}{l|rr|rrrrr|rrrrr}
\toprule
\multicolumn{1}{l}{Motivation/Sector} & Total & Geo. & Top-1 & Top-2 & Top-3 & Top-4 & Top-5 & Avg. & Std. & Q1 & Q2 & Q3 \\

\midrule

Financial / Cybercrime & 6513 & 120 & Costa Rica (89.5\%) & Brazil (85.1\%) & Peru (83.3\%) & Chile (82.5\%) & Spain (80.8\%) & 36.8 & 22.8 & 19.0 & 33.2 & 52.3 \\

Espionage \& Intelligence Collection & 4015 & 121 & North Korea (100.0\%) & Brunei (100.0\%) & Cambodia (98.2\%) & Tajikistan (96.3\%) & Libya (95.8\%) & 62.3 & 22.4 & 45.7 & 65.1 & 78.0 \\

Sabotage & 279 & 50 & \textbf{Ukraine} (18.1\%) & \textbf{Albania} (14.3\%) & \textbf{Saudi Arabia} (11.2\%) & \textbf{Latvia} (9.1\%) & \textbf{Iran} (9.0\%) & 3.2 & 3.7 & 1.0 & 1.9 & 3.6 \\

Political \& Ideological Influence & 269 & 82 & \textbf{Latvia} (36.4\%) & \textbf{Estonia} (34.8\%) & \textbf{Lithuania} (33.3\%) & \textbf{Albania} (33.3\%) & \textbf{Finland} (25.0\%) & 7.7 & 8.2 & 2.3 & 4.2 & 9.9 \\

Malware Propagation & 252 & 32 & \textbf{Tunisia} (18.2\%) & Algeria (4.8\%) & Ecuador (3.7\%) & Morocco (3.4\%) & Argentina (3.0\%) & 1.9 & 3.2 & 0.5 & 1.2 & 2.3 \\

Military \& Cyber Warfare & 98 & 46 & \textbf{Ukraine} (10.6\%) & \textbf{Central Asia} (9.3\%) & \textbf{Latvia} (9.1\%) & \textbf{Lithuania} (8.3\%) & \textbf{Georgia} (6.7\%) & 2.5 & 2.7 & 0.6 & 1.3 & 3.0 \\

Reconnaissance \& Security Testing & 20 & 3 & Israel (0.4\%) & Japan (0.2\%) & USA (0.1\%) &  &  & 0.2 & 0.2 & 0.2 & 0.2 & 0.3 \\

Personal & 6 & 1 & China (0.2\%) &  &  &  &  & 0.2 & 0.0 & 0.2 & 0.2 & 0.2 \\

Strategic Objectives & 4 & 3 & China (0.2\%) & South Korea (0.2\%) & USA (0.1\%) &  &  & 0.2 & 0.1 & 0.1 & 0.2 & 0.2 \\

Illicit Content \& Perversion & 2 & 1 & USA (0.1\%) &  &  &  &  & 0.1 & 0.0 & 0.1 & 0.1 & 0.1 \\
\midrule

Government \& Public Administration & 2493 & 121 & NATO (100.0\%) & Uzbekistan (91.7\%) & Laos (90.6\%) & Guatemala (90.0\%) & Brunei (90.0\%) & 55.9 & 19.2 & 40.9 & 52.7 & 71.4 \\

\makecell[l]{Information Technology, Telecommunications \\ \& Cybersecurity} & 1934 & 119 & \textbf{Tunisia} (54.5\%) & Albania (52.4\%) & Cyprus (52.0\%) & South Africa (48.1\%) & Asia-Pacific (45.5\%) & 25.7 & 10.2 & 18.2 & 24.2 & 32.3 \\

Finance, Banking \& Insurance & 1663 & 120 & \textbf{Uruguay} (66.7\%) & Peru (59.5\%) & Slovenia (56.2\%) & New Zealand (54.5\%) & Panama (54.5\%) & 26.4 & 13.9 & 15.8 & 24.2 & 34.3 \\

Individuals \& Personal Services & 1651 & 111 & \textbf{Morocco} (41.4\%) & \textbf{Libya} (33.3\%) & \textbf{Syria} (31.0\%) & Algeria (28.6\%) & Iran (27.6\%) & 12.7 & 7.2 & 7.4 & 11.8 & 16.3 \\

Military & 820 & 120 & \textbf{Venezuela} (55.0\%) & \textbf{Latvia} (48.5\%) & \textbf{Sri Lanka} (48.0\%) & \textbf{NATO} (47.4\%) & \makecell[r]{Bosnia and\\Herzegovina (42.9\%)}  & 19.0 & 11.0 & 11.3 & 16.7 & 25.1 \\

Energy, Utilities \& Natural Resources & 711 & 117 & \textbf{Qatar} (33.3\%) & South America (30.0\%) & Venezuela (30.0\%) & Saudi Arabia (30.0\%) & Guatemala (30.0\%) & 14.6 & 6.5 & 9.5 & 13.6 & 18.4 \\

Manufacturing, Industrial \& Materials & 705 & 110 & \textbf{East Asia} (35.5\%) & \textbf{South America} (30.0\%) & \textbf{Taiwan} (28.3\%) & Asia-Pacific (27.3\%) & Japan (23.1\%) & 11.2 & 6.4 & 6.3 & 10.2 & 14.9 \\

Education \& Research & 702 & 117 & \textbf{Cyprus} (36.0\%) & \textbf{Serbia} (29.2\%) & \textbf{Algeria} (28.6\%) & \textbf{Asia-Pacific} (27.3\%) & \textbf{Tunisia} (27.3\%) & 14.5 & 6.4 & 10.5 & 14.3 & 17.2 \\

Healthcare \& Life Sciences & 610 & 112 & \textbf{Ireland} (38.7\%) & Costa Rica (21.1\%) & Chile (21.1\%) & Brunei (20.0\%) & United Kingdom (19.1\%) & 10.0 & 5.6 & 5.9 & 9.3 & 13.3 \\

Retail, Commerce \& Consumer Services & 470 & 108 & \textbf{Panama} (27.3\%) & \textbf{Asia-Pacific} (22.7\%) & \textbf{South America} (20.0\%) & \textbf{South Africa} (19.2\%) & Ireland (16.1\%) & 6.8 & 4.6 & 3.5 & 5.7 & 9.1 \\

Media, Advertising \& Journalism & 446 & 113 & \textbf{Brunei} (50.0\%) & \textbf{Yemen} (42.9\%) & \textbf{Tunisia} (36.4\%) & \textbf{Palestine} (33.8\%) & \textbf{Lebanon} (31.5\%) & 12.6 & 8.4 & 7.1 & 10.0 & 16.0 \\

NGOs, Advocacy \& Civil Society & 411 & 117 & \makecell[r]{\textbf{Bosnia and} \\\textbf{Herzegovina} (42.9\%)} & \textbf{Tunisia} (36.4\%) & \textbf{North Korea} (35.7\%) & \textbf{Morocco} (34.5\%) & \textbf{Yemen} (28.6\%) & 11.2 & 7.5 & 6.2 & 9.1 & 12.1 \\

Cryptocurrency \& Fintech & 392 & 86 & \textbf{Slovenia} (25.0\%) & \textbf{Nigeria} (21.1\%) & \textbf{Uruguay} (16.7\%) & \textbf{Costa Rica} (15.8\%) & \textbf{Kenya} (14.3\%) & 6.0 & 4.5 & 2.9 & 5.0 & 7.4 \\

Transportation, Logistics \& Maritime & 350 & 108 & \textbf{Denmark} (27.3\%) & \textbf{Kuwait} (24.0\%) & \textbf{Estonia} (21.7\%) & \textbf{Finland} (20.0\%) & Lithuania (18.8\%) & 9.2 & 4.6 & 5.9 & 8.3 & 11.5 \\

Aerospace, Aviation \& Space & 309 & 108 & \textbf{Albania} (33.3\%) & \textbf{Mongolia} (16.9\%) & \textbf{Denmark} (15.2\%) & \textbf{Finland} (15.0\%) & \textbf{Sweden} (14.8\%) & 7.0 & 4.2 & 4.4 & 6.2 & 8.5 \\

Critical Infrastructure \& ICS/OT & 257 & 77 & \textbf{Kuwait} (10.0\%) & \textbf{Sudan} (10.0\%) & \textbf{Central Asia} (9.3\%) & \textbf{Denmark} (9.1\%) & Ukraine (8.5\%) & 3.7 & 2.3 & 2.0 & 3.1 & 4.8 \\

Entertainment, Sports \& Gambling & 197 & 80 & \textbf{Thailand} (11.5\%) & \textbf{Hong Kong} (11.3\%) & Nigeria (10.5\%) & Brunei (10.0\%) & Philippines (10.0\%) & 4.1 & 2.7 & 1.9 & 3.4 & 5.6 \\

Hospitality \& Travel & 148 & 86 & \textbf{Brunei} (20.0\%) & \textbf{Bahrain} (15.4\%) & \textbf{Latin America} (12.1\%) & \textbf{Oman} (11.1\%) & \textbf{Guatemala} (10.0\%) & 4.4 & 3.1 & 2.6 & 3.6 & 5.5 \\

Legal \& Judiciary & 117 & 84 & \textbf{Slovenia} (18.8\%) & \textbf{Panama} (18.2\%) & \textbf{Cyprus} (16.0\%) & \textbf{Norway} (12.8\%) & Algeria (9.5\%) & 4.2 & 3.6 & 1.9 & 3.3 & 5.3 \\

Real Estate, Construction \& Engineering & 115 & 71 & \textbf{Slovenia} (12.5\%) & \textbf{New Zealand} (12.1\%) & \textbf{Venezuela} (10.0\%) & \textbf{Algeria} (9.5\%) & Switzerland (7.4\%) & 3.5 & 2.4 & 2.0 & 2.8 & 4.4 \\

Consulting \& Professional Services & 114 & 55 & \textbf{Brunei} (10.0\%) & \textbf{Panama} (9.1\%) & \textbf{Tunisia} (9.1\%) & Norway (7.7\%) & Ireland (6.5\%) & 3.0 & 2.3 & 1.2 & 2.6 & 4.0 \\

Food \& Agriculture & 74 & 61 & \textbf{Costa Rica} (15.8\%) & \textbf{Chile} (10.5\%) & \textbf{South America} (10.0\%) & \textbf{Panama} (9.1\%) & \textbf{Cyprus} (8.0\%) & 2.9 & 2.8 & 1.2 & 2.2 & 3.0 \\

Religion \& Culture & 28 & 35 & \makecell[r]{\textbf{Bosnia and} \\\textbf{Herzegovina} (35.7\%)} & \textbf{Cambodia} (12.5\%) & \textbf{Indonesia} (7.9\%) & \textbf{East Asia} (6.5\%) & Malaysia (5.5\%) & 3.2 & 6.2 & 0.9 & 1.2 & 2.9 \\
\bottomrule

\end{tabular}
}
\end{table*}

\end{document}